\documentclass[a4,aps,prb,amsmath,amssymb,superscriptaddress, showpacs, twoside,epsfig]{revtex4}

\usepackage{graphicx}
\usepackage{epstopdf} 
\newcommand{\be}[1]{ \begin{eqnarray} \mbox{$\label{#1}$} }

\newcommand{\ee}{\end{eqnarray}}
\newcommand{\pref}[1]{(\ref{#1})}

\newcounter{mycount}

\newcommand\ie {{\it i.e. }}
\newcommand\eg {{\it e.g. }}
\newcommand\etc{{\it etc. }}

\newcommand\cf {{\it cf.  }}

\newcommand\half{\frac 1 2 }

\newcommand\ket [1] {|#1 \rangle }
\newcommand\bra [1] {\langle #1 |}

\newcommand{\av}[1]{\langle #1\rangle}

\newcommand\noi{\noindent}
\newcommand{\bi}{\begin{itemize}}
\newcommand{\ei}{\end{itemize}}
\newcommand{\ba}{\begin{eqnarray}}
\newcommand{\ea}{\end{eqnarray}}
\newcommand{\ds}{\displaystyle}

\newcommand{\bfr}{\mathbf{r}}

\usepackage{lscape}

\begin{document}

\title{Composite fermion wave functions as conformal field theory correlators}
\author{T.H. Hansson} 
\affiliation {Department of Physics, Stockholm University,
AlbaNova University Center,
SE - 106 91 Stockholm, Sweden} 
\author{C,-C. Chang}
\author{J.K. Jain}
\affiliation{Physics Department, 104 Davey Lab, The Pennsylvania State University, 
University Park, Pennsylvania 16802}
\author{S. Viefers}
\affiliation{ Department of Physics, University of Oslo, P.O. Box 1048 Blindern, 0316 Oslo, Norway}

\date{\today}

\begin{abstract} 
It is known that a  subset of fractional quantum Hall wave functions has been expressed as 
conformal field theory (CFT) correlators, notably the Laughlin wave function at filling 
factor $\nu=1/m$ ($m$ odd) and its quasiholes, and the Pfaffian 
wave function at $\nu=1/2$ and its quasiholes. 
We develop a general scheme for constructing composite-fermion (CF) wave functions
from conformal field theory.  Quasiparticles at  
$\nu=1/m$ are created by inserting anyonic 
vertex operators, $P_{\frac{1}{m}}(z)$, that replace a subset of the electron operators
in the correlator. 
The one-quasiparticle wave function is identical to the corresponding CF wave 
function, and the two-quasiparticle wave function has correct fractional charge and statistics and is numerically almost identical to the corresponding CF wave function.  
We further show how to  exactly represent the CF wavefunctions in the Jain series $\nu = s/(2sp+1)$
 as the CFT correlators  of a new type of  fermionic vertex 
operators, $V_{p,n}(z)$, constructed from $n$ free compactified 
bosons; these operators provide the CFT representation of composite fermions
carrying $2p$ flux quanta in the $n^{\rm th}$ CF Landau level.  We also construct the corresponding 
quasiparticle- and quasihole operators and argue that they have the expected fractional charge and statistics. 
For filling fractions 2/5 and 3/7 we show that the chiral CFTs that describe the bulk wave functions are identical to those 
given by Wen's general classification of quantum 
Hall states in terms of $K$-matrices and $l$- and $t$-vectors, and 
we propose that to be generally true. 
Our results suggest a general procedure for constructing quasiparticle wave functions for other fractional Hall states, as well as for constructing ground states at filling fractions not contained in the principal Jain series.  
\end{abstract}

\maketitle

\newcommand\jas[3]{(z_{#1} - z_{#2})^{#3}}
\newcommand\prs[2] {\prod_{#1}\!^{(#2)}}
\newcommand\pr[3] {\prod^{(#3)}_{#1<#2}\!}
\newcommand\prjas[4] {\pr #1 #2 {#3}  \jas #1 #2 #4}
\newcommand\vmea[1] {e^{i\sqrt{m} \varphi_1 (z_#1)}} 
\newcommand\vtmea[1] {\partial_{z_#1} e^{i(\sqrt{m}-\frac 1 {\sqrt m}) \varphi_1 (z)}}
\newcommand\vme {e^{i\sqrt{m} \varphi_1 (z)}} 
\newcommand\vtme {\partial e^{i(\sqrt{m}-\frac 1 {\sqrt m}) \varphi_1 (z)}}
\newcommand\vm[1] {V_1(z_{#1}) }
\newcommand\vmn[1] {V_1(z_{#1}) }
\newcommand\bvm[1] {V_1(\overline z_{#1}) }
\newcommand\vtm[1] { P_{\frac{1}{m}}(z_{#1}) }
\newcommand\bhvtm[1] { \hat P_{\frac{1}{m}}(\bar z_{#1}) }
\newcommand\hvtm[1] { \hat P_{\frac{1}{m}}(z_{#1}) }
\newcommand\bvtm[1] { P_{\frac{1}{m}}(\bar z_{#1}) }
\newcommand\holee[1]{e^{ \frac i {\sqrt m} \varphi_1 (\eta_{#1} ) } }
\newcommand\hole[1] {H_{\frac{1}{m}}(\eta_#1)}

\newcommand\zij{z_{ij}}
\newcommand\bzij{Z_{ij}}
\newcommand\eb{\bar\eta}
\newcommand\beb{\bar N}
\newcommand{\epl}{\eta_+}
\newcommand{\emi}{\eta_-}

\newcommand\veps{\varepsilon}
\newcommand\cL{{\cal L}}
\newcommand\La{K}
\newcommand\btt{{\bf  t}}
\newcommand\bll{{\bf l}}

\section{Introduction}
\newcommand{\CFT } {conformal field theory } 
\vskip 3mm\noi
The evidence for an intriguing  
connection between \CFT (CFT) and the fractional quantum Hall effect (FQHE) was accumulating in the 1980s.  
It was realized that the effective low-energy theory of the FQHE is a topological field theory of the Chern-Simons type, where the exchange phases of the anyonic quasiparticles and quasiholes are coded in the braiding properties of the corresponding Wilson loops\cite{CStheory}.  Witten's subsequent demonstration
that the braiding of the Wilson loops are reflected in the  correlation functions 
of certain CFTs\cite{witten} suggested  a CFT-FQHE relationship, 
which was further strengthened by Wen, who proposed that the gapless chiral edge modes of a FQH-droplet are described by a chiral $1+1$ dimensional CFT\cite{wen}.  
It was also noticed that the holomorphic part of the Laughlin wave function takes the form of a correlator of bosonic exponents, or vertex operators, in a two dimensional CFT\cite{MR,Fubini}.

The 1991 paper by  Moore and Read was particularly important since it synthesized  many of these ideas and made an explicit conjecture about the CFT description of 
quantum Hall (QH) states containing two parts:  1. ``Representative" electronic wave functions for the ground state and its quasiparticle and quasihole excitations are correlation functions, or, more precisely, conformal blocks, in a rational \CFT (RCFT) where the various particles correspond to different primary fields. 2. The very same RCFT describes the edge excitations of the corresponding FQH droplet. 
In their paper Moore and Read gave some striking  
circumstantial arguments to support their conjecture, and they also showed that many 
FQH states, namely the Laughlin state, the states in the Halperin-Haldane hierarchy, their  quasihole excitations, the Halperin spin singlet state\cite{halperin}, and the Haldane-Rezayi spin singlet pairing state\cite{rezhal}, may be represented in terms of conformal blocks. 
All this might have been criticized for being just a reformulation of old results, but Moore and Read also used the CFT formalism to propose a new $\nu = 1/2$ state, 
the so-called Pfaffian wave function, which is tentatively assigned to the observed $\nu = 5/2$ FQHE. The quasiholes in this state have charge $q = 1/4$ rather than $q = 1/2$ expected from the filling fraction, and exhibit non-Abelian fractional statistics. To establish the latter it was essential to use CFT technology.\footnote{
 Later the non-Abelian statistics has also been understood in the context of d-wave paired superconductors\cite{pwave} and has also been studied numerically\cite{Tser}.  }

Despite this advance one and a half decades ago,  the program of establishing a 
one-to-one correspondence 
between QH states and conformal field theory has remained incomplete.  
No explicit \CFT expressions have so far been established for many important FQHE states; in particular, despite interesting progress\cite{flohr},
this is the case for the ground state wave
functions of the prominent FQHE series $\nu=s/(2sp\pm1)$, and their related quasihole or quasiparticle excitations. (Expressions for the states in the Haldane-Halperin hierarchy were given in Ref. \onlinecite{MR}, but these are indirect, involving multiple integrals over auxillary quasihole coordinates.)
Surprisingly, a proper \CFT representation does not exist even for the 
quasiparticles -- as opposed to quasiholes -- of the FQHE state at $\nu=1/m$
and the Pfaffian wave function at $\nu=1/2$.
\footnote{
An asymmetry in the description of the quasiparticles and quasiholes at 
$\nu=1/m$ has been a striking feature of other descriptions as well. 
In Laughlin's theory, a quasihole at position $\eta$ is  
represented as a vortex $\prod_i (z_i - \eta)$, while a quasiparticle is created 
by a complicated  operator involving many derivatives. 
The fractional statistics of the quasiholes is easy to derive, while the statistics of the quasiparticles eludes a precise analytical treatment. 
In the Ginzburg-Landau-Chern-Simons effective theories, the quasiholes and quasiparticles are described by vortices and anti-vortices respectively, and again there is an asymmetry in the description.\cite{tafel} }

It is worth reminding ourselves what we
can hope to accomplish using CFT techniques:
We cannot ``derive" the FQHE wave functions, since the CFT 
does not contain any information about the actual interelectron interaction.
It is true that the short distance behavior of the electronic wave functions is reflected in the operator product
expansion of the pertinent CFT vertex operators, but only in the simplest cases can this be directly
related to a potential of the Haldane-Kivelson-Trugman type. 
Thus we can only hope to get "representative wave functions" in the sense of Moore and Read, and any new candidate wave function 
suggested by the CFT approach must be tested and confirmed against exact solutions 
of the Schr\"odinger equation known for small systems. 
The crucial question is if the CFT wave functions are sufficiently natural and simple 
to give new insight into the 
physics of the problem, facilitate computations of quantities like local charge 
and braiding statistics, and most importantly,  inspire new generalizations.  
Finally, we should point out that we know of no 
general microscopic principle that requires that the correlated 
quantum mechanical wave functions of interacting electrons in the lowest Landau level 
should be expressible as simple correlation functions of certain vertex operators 
in a two dimensional Euclidean rational conformal field theory.

An insight into the general FQHE states comes from the composite fermion (CF) 
formalism\cite{Jain,review}. 
Here the experimentally prominent Jain states at $\nu = s/(2sp +1)$ are formed from  $s$ filled Landau levels of ``composite fermions," which are electrons carrying $2p$ flux quanta. Other CF states, as \eg the Pfaffian, which is the preferred candidate for the observed $\nu = 5/2$ state, can be formed by various BCS type pairing mechanisms\cite{MR,pwave}. In the CF description, a quasihole is obtained simply by removing a composite fermion from an incompressible FQHE state, and a quasiparticle  is a 
composite fermion in a higher, otherwise empty CF Landau level (LL).  (CF Landau levels are also called $\Lambda$ levels.)  Explicit wave 
functions are constructed for all ground states and their quasiparticle and quasihole 
excitations.  (The asymmetry between quasiparticles and quasiholes occurs 
since they reside in different CF Landau levels.) 
The CF approach is very successful, both in comparison with experiments and with numerical studies of two-dimensional electron gases in strong magnetic fields\cite{review}.

The issue of fractional charge and fractional statistics of the composite fermions is a subtle one. 
The quasiparticles and quasiholes are composite fermions added to or removed from a 
CF Landau level.  From one perspective, they have unit charge and fermionic statistics. 
Indeed, the addition of one composite fermion increases the number of electrons, and 
hence the net charge, by one unit, and the fermionic statistics of 
composite fermions has been confirmed by numerous experiments 
(\eg the observation of their Fermi sea).  On the other hand, the CF quasiparticles 
and quasiholes have a fractional ``local charge" 
(where the local charge is the charge measured relative to the background FQHE state) and a fractional braiding 
statistics \cite{Leinaas,kjons,jain2,review}.  These properties capture the physics
 that adding or removing a composite fermion causes nonlocal changes 
in the state, because the vortex, a constituent of the 
composite fermion, is a nonlocal object.  This should be contrasted with the 
analogous process in the  integral QHE, which is essentially 
local  (the Landau level projection destroys locality only on the scale of the magnetic length $\ell$), and  can be described by a local, charge-$e$ operator  $\psi_\alpha^\dagger(\vec x)$, where the subscript denotes the Landau level index. 
No such local operator can be constructed for the creation 
of a composite fermion, since the local charge of the quasiparticle differs from that of the electron.  The fractional statistics of the quasiparticles also implies that 
they cannot be described by local operators, as emphasized by Fr{\"o}hlich and Marchetti\cite{froh}.
Even though fractional charge and fractional statistics cannot be 
read off directly from the CF wave functions, they nonetheless contain 
that information, not surprising in view of the fact that the CF construction provides a good description 
of all the low energy states.  We mention here the quasiparticles at $\nu=1/m$, for which 
the CF wave function differs from that proposed earlier by Laughlin\cite{laughlin83}. 
The calculation of the 
Berry phase associated with two-CF quasiparticle exchange, originally performed   
by Kj\o nsberg and Leinaas\cite{kjons} and subsequently by Jeon and collaborators\cite{jain2},  shows that the braiding statistics for the CF quasiparticles has a sharply defined
fractional value;  for the Laughlin quasiparticles, in contrast, 
numerical calculations do not produce a convergent result for the statistical angle\cite{kjons2}.

In this paper we  establish a firm 
connection between CF wave functions and CFT correlators.  Specifically: 

\begin{enumerate}

\item 
We construct the quasiparticles of $\nu=1/m$ ($m$ odd) 
using a new kind of anyonic vertex operators $P_{\frac{1}{m}}$. For a single quasiparticle,
the resulting wave function is identical to that obtained using the CF theory. A 
generalization to two or more quasiparticles produces wave functions that are very similar to the CF wave functions but not identical.  For two quasiparticles at $\nu=1/3$, the overlap between the two wave functions is typically 99.99\% for as many as 40 electrons.  

\item 
We show that the ground state wave functions in the Jain series $\nu = n/(2np+1)$ are exactly given by 
sums of CFT correlators of a set of vertex 
operators, $V_{np}$,  which in the CF language correspond to creating composite fermions in higher CF Landau levels. 
 
\item 
We generalize the construction of the quasiparticle operator $P_{\frac{1}{m}}$, as well
as of the quasihole operators, to higher levels in the Jain sequence; at
level $n$, there are $n$ independent hole operators and one quasiparticle operator. 
The vertex operator $V_{n,p}$ at level $n$ is closely related to the quasiparticle 
operator at level $n-1$.

\item We demonstrate that the very CFT that yields the CF wave functions also  
directly defines an edge theory for the Jain states that is precisely the one expected from the general arguments given by Wen\cite{wen}. 

\end{enumerate}

Our CFT construction has many advantages. (i)  At the technical level, it produces accurate wave 
functions directly in the lowest Landau level with no need for projection, and the charge and statistics of the quasiparticles are revealed in the algebraic properties of the corresponding operators, just as in the case of the quasiholes of the $\nu=1/m$ states. 
(ii) Although the effective edge theory for the Jain states was known from general principles, we provide a direct derivation from a CFT where the conformal 
blocks yield microscopically accurate bulk wave functions.  
(iii) It gives a new insight and suggests new extensions; 
a generalization of this work produces natural ans{\"a}tze for quasiparticle wave functions for more complicated CF states such as the Moore-Read Pfaffian state, as well as for ground states at fractions ({\em e.g.}, 4/11), which do not belong to the principal Jain series.  

The paper is organized as follows. 
In the next section we explain the basic ideas behind our construction and give explicit wave functions for one- and two-quasiparticles, as well as that for a quasiparticle-quasihole pair. The  general structure of the CFT description of the states in the Jain series is discussed in section \ref{sec:III}, while the detailed technical proof for the equivalence between the CF and the CFT wave functions is left for Appendix \ref{app:B}.
In section \ref{sec:IV} we explain the construction of the edge theory,  and in section \ref{sec:V} we construct localized quasiparticle states and show how to extract charge and statistics from the relevant Berry phases; the latter can be ascertained analytically if we make a random phase assumption. Some details of the calculations are found in Appendix \ref{app:C}. 
Section \ref{sec:V} presents numerical calculations supporting our claims in sections \ref{sec:II} and \ref{sec:V} and, finally, a summary is found in section \ref{sec:VII}.  A short report on parts of this work has been published previously\cite{HanssonI}.

\section{One and two quasiparticles in the Laughlin state} \label{sec:II}
\subsection{The ground state and the quasihole states}
We first review some of the basic formalism of the CFT construction of QHE wave functions, in particular
the construction of the ground state and quasihole  wave functions at the Laughlin fractions $\nu =  1/m$, where $m$
is an odd integer.
Following Moore and Read\cite{MR}, we introduce the normal-ordered vertex operators,
\be{vo}
V_1(z) &=& : \vme : \\
H_{\frac{1}{m}}(\eta) &=& :e^{ \frac i {\sqrt m} \varphi_1 (\eta) } : \label{hole} \, ,
\ee
where the normal ordering symbol $:\ \ :$, will be suppressed in the following. 
The free massless boson field,    $\varphi_1$, is    normalized so as to 
have the (holomorphic) two point function
\be{twop}
\av{ \varphi_1 (z) \varphi_1 (w) } = - \ln (z - w)   \, ,
\ee
so that the the vertex operators obey the relations 
\be{verrel}
e^{i\alpha \varphi_1(z)}  e^{i\beta \varphi_1(w)}
	 &=& e^{i\pi \alpha\beta}e^{i\beta \varphi_1(w)} e^{i\alpha \varphi_1(z)} 
	= (z-w)^{\alpha\beta} e^{i\alpha \varphi_1(z) + i\beta \varphi_1(w)} \nonumber \\
	&\sim &(z-w)^{\alpha\beta} e^{i(\alpha  + \beta) \varphi_1(w)} 
\ee	 
where the last line expresses the operator product expansion (OPE) in the limit $z\rightarrow w$. 
From \pref{verrel} follows 
$V_1(z) V_1(w) + V_1(w) V_1(z) = 0$, and 
$H_{\frac{1}{m}}(z) H_{\frac{1}{m}}(w)- e^{i\pi/m}H_{\frac{1}{m}}(w) H_{\frac{1}{m}}(z) =0$. 
The first of these reflects that the electrons are fermions, while the second is
appropriate for fractional statistics as discussed in reference [\onlinecite{MR}].

 We normalize the (holomorphic) $U(1)$ charge density operator as
\be{cdo}
 J(z) =\frac i {\sqrt m} \partial_z \varphi_1 (z)   \,
\ee
so the corresponding charge is given by 
\be{co}
{\cal Q} = \frac 1 {\sqrt m} \frac 1 {2\pi } \oint dz\,  \partial_z \varphi_1 (z) ,
\ee
where the contour encircles the whole system. 
The $U(1)$ charges, $Q=1$ of the electron and $Q=1/m$ of the quasihole, can be read directly from the commutators 
$[{\cal Q}, V_1(z) ] = V_1(z) $ and $ [{\cal Q}, H_{\frac{1}{m}}(\eta) ] = 
\frac 1 m H_{\frac{1}{m}}(\eta)$. 
It is noted that $Q$ does not give the electric charge; rather it has the interpretation of vorticity as seen from \pref{verrel}.
Introducing a positive vorticity in a homogenous state  corresponds to a local depletion of the electron liquid, while a  negative vorticity amounts to a local increase in density. Thus the excess electron number  compared with the ground state created by an operator with $U(1)$ charge $Q$ is given by 
\be{elnr}
\Delta n = \delta n - Q,
\ee
where the integer $\delta n$ is the number of electrons added by the operator. If the argument of the operator is an electron coordinate, $z_i$, one electron is added, while no electron is added if the argument is a quasihole coordinate $\eta_i$.  (The idea of 
binding of an electron and $m$ vortices was implicit in Laughlin's original work, and was made explicit  by Halperin \cite{halperin}, Girvin and MacDonald \cite{GM} and Read \cite{read}.)  

The total electric charge of a particle  is given by $Q_{el} = -e\Delta n = e(Q - \delta n)$. Note that the excess charge associated with the addition 
of an electron is zero, as expected, because 
this expands the droplet without creating any local charge variation. 

The (un-normalized)  $\nu = 1/m$ Laughlin wave function can now be written as (for notational convenience, we write $\Psi (z_i)$ instead of $\Psi(\{z_i\})$):
\be{La}
\Psi_L (z_i) 
 &= &\bra 0   {\cal R}\{\vm 1 \vm 2 \dots \vm {N-1} \vm N  e^{-i\sqrt m \rho_m \int d^2z\, \varphi_1(z) } \} \ket 0  \\
& \equiv      &         \av{ \vm 1 \vm 2 \dots \vm {N-1} \vm N }_{1/m} \nonumber \\
&= & \prod_{i<j} \jas i j m e^{-\sum_i  |z_i|^2/4\ell^2} ,\nonumber
\ee
where ${\cal R}$ denotes radial ordering. 
The second line defines the average $\av{\dots}_{1/m}$, and the third follows for the 
ordering $|z_1| \ge |z_2| \ge \dots |z_N|$,  which will be assumed below unless indicated otherwise. 
In the following, we shall suppress the subscript $1/m$ whenever it is clear to what ground state we are referring. 
The exponential operator in \pref{La} corresponds to a constant background particle density, $\rho_m = -\rho_0/m$, where $\rho_0= 1/2\pi\ell^2$ is the density of a filled Landau level. This is 
necessary since the $U(1)$ 
charge neutrality condition, known from the Coulomb gas formulation, in the  CFT ensures that the correlator vanishes unless $N = \rho_m \int d^2z = \rho_m A$, which defines the area, $A$,  of the system. As explained in reference  [\onlinecite{MR}], the background charge will produce the correct gaussian factor $e^{-\sum_i^N |z_i|^2/4\ell^2}$ characteristic of the lowest Landau level wave function. For a more detailed discussion of this background charge prescription, see Appendix \ref{app:A}. 

The wave function for a collection of Laughlin quasiholes is also easily written:
\be{lqh}
\Psi_L (\eta_1\dots \eta_n ; z_i) =  \av{ \hole 1 \hole 2 \dots \hole n \vm 1 \vm 2 \dots \vm {N-1} \vm N } .
\ee
In this case the charge neutrality condition reads $N + n/m = \rho_m A'$, indicating an expansion of the droplet.  
From the general relation \pref{verrel}	 
we get $H_{\frac{1}{m}}(z) V_1(w) +V_1(w) H_{\frac{1}{m}}(z) =0$ 
 which guarantees that \pref{lqh} is uniquely defined and analytic in the electron coordinates. 

Very little of the rather sophisticated mathematics of CFT will be used in this paper, but a few formal comments are in order. A CFT is in general not defined by a Lagrangian, but by an operator product  algebra, or set of fusion rules, together with a specification of the field content defined by the so-called primary fields. The CFTs of interest here are defined by a Lagrangian describing a collection of free bosons, $\varphi_i$, compactified on circles of radius $R_i = \sqrt{m_i}$ where $m_i$ are odd integers.   
The primary fields are given by the chiral vertex operators $V(z) = e^{ i\sum_i \frac {q_i} {R_i }\varphi_i (z)}$ where the integers $q_i$ define the charge lattice describing the possible ``electric"  charges in the Coulomb gas formulation of the CFT. The vertex operators satisfy an extended chiral algebra that,  together with the charge lattice, defines the relevant CFT, which in this case is called a ``rational torus" with radii $\sqrt m_i$;
this  is an example of a rational CFT. Acting on the primary fields with the generators of the conformal group gives families of  ``descendant fields", which can be expressed using derivatives of  the parent primary fields. Such descendant fields will be important in the construction of quasiparticle operators presented in the next section. 
The full CFT contains fields of both chiralities and has correlation functions that can be written as (in general a sum over) products of holomorphic and anti-holomorphic factors, so-called conformal blocks. The holomorphic blocks are precisely the correlation functions of chiral vertex operators that we have identified with the electronic wave functions. In general, these blocks also depend parametrically on quasiparticle and quasihole coordinates, and acquire nontrivial phase factors, called monodromies, when these coordinates are transported along closed loops. It is these monodromies that reproduce the braiding phases that also can be calculated from the expectation values of Wilson loops in a Chern-Simons theory.  
A detailed discussion of the conditions that a CFT has to fulfill in order to describe a QH state can be found in Ref. \onlinecite{frohlich}. 

\subsection{One quasiparticle}

The most immediate guess\cite{MR} for a quasiparticle operator would be to simply change the sign in the exponent in the quasihole operator of  \pref{hole}, \ie to use $e^{-\frac i {\sqrt m} \varphi_1(\eta)}$.  That, however, introduces unacceptable singular terms $\sim \prod_i (z_i - \eta)^{-1}$ in the electronic wave function. Inspired by the CF wave functions, we instead define a quasiparticle operator, $ P_{\frac{1}{m}}(z)$, which has 
a $U(1)$ charge  $(1- 1/m)$, and that  will {\em replace} one of the the original electron operators $V_1(z)$. We can
thus think of $P(z)$ as a modified electron operator, but with a different amount of vorticity. 
 The excess electric charge associated with such a modification is the difference between the charges of the operators $V_1$ and $P_{\frac{1}{m}}$ \ie    $\Delta Q_{el} =  e((1-1/m)-1) = -e/m$, as appropriate for a quasiparticle
at $\nu=1/m$. The modified electron operator is given by
\be{qpo}
 P_{\frac{1}{m}}(z) = \vtme ,
\ee
and the wave function for a single quasiparticle with angular momentum $l$ is 
written as 
\be{1qp}
\Psi_{1qp}^{(l)}  (z_i) &=&  {\cal A}\{  z_1^l   e^{- |z_1|^2 /4m\ell^2}  \av{ \vtm 1 \vm 2  \dots  \dots    \vm N     } \}\\ 
&=&\sum_i(-1)^{i+1} z_i^l \, e^{- |z_i|^2 /4m\ell^2}  \av{ \vtm i \prod_{j\ne i} \vm j   }  \nonumber \\
&=&\sum_i (-1)^i z_i^l \pr j k {i} (z_j-z_k)^m 
\partial_i  \prod_{l \neq i}(z_l-z_i)^{m-1}, \nonumber
\ee
where ${\cal A}$ denotes anti-symmetrization of the coordinates.  The second line follows by noting that the anti-symmetrized product has the form of a Slater determinant which is then expanded by the first row. From \pref{verrel} we get 
$P_{\frac{1}{m}}(z) V_1(w)- V_1(w) P_{\frac{1}{m}}(z) =0$, so the radial reordering of the quasiparticle operator does not give rise to any sign change. 
The anti-symmetrization with respect to the remaining coordinates is trivial  since $V_1(z) V_1(w) + V_1(w) V_1(z) = 0$. 
The charge neutrality condition now reads $N-1+(1-\frac 1 m) = \rho_m A''$, so the droplet has 
undergone a small contraction, as expected for a quasiparticle. 

While the exponent of \pref{qpo} follows naturally from the above charge requirement (and may be viewed as a combination of an electron operator and an ``inverse" quasihole operator), the
derivative has been put in ``by hand".  Without the derivative, the wave function \pref{1qp} can be shown to be identically zero. Technically,  $P_{\frac{1}{m}}(z)$ is 
a descendant of the primary field, $e^{i(\sqrt{m}-\frac 1 {\sqrt m}) \varphi (z)}$, 
a construction that naturally generalizes to more complicated QH states \cite{HHV}. 
Note that the derivative in \pref{1qp} acts only on the holomorphic part of the wave function. \footnote
{A more careful evaluation of the correlators  using a regularized background charge (\cf Appendix A) does give  a contribution also from the exponential. In order to have holomorphic wave function this must be cancelled by replacing $\partial_i$ with a suitable covariant derivative. Since this in the end amounts to a mere change of notation, we simply use the rule that the derivatives do not act on the exponential part in the correlation function.}

The quasiparticle wave function of 
 \pref{1qp} has a different character than those written 
above for the ground and the quasihole states, in that it is a sum over correlators,
 and that it 
involves prefactors $f_1(z_i) = z_i^l   e^{- |z_i|^2 /4m\ell^2}$.  
The factor $z_i^l$ sets the angular momentum, while the exponential factor is chosen 
to give the correct lowest Landau level (LLL)  electronic wave function: 
Due to its modified charge, the quasiparticle 
operator $P_{\frac{1}{m}}(z_i)$ gives rise to an exponential factor $\exp(- |z_i|^2(1-1/m) /4\ell^2)$, and the
compensating prefactor ensures that the overall gaussian factor is $\exp\{- \sum_j |z_j|^2 /(4\ell^2)  \}$. Here and in the following, we suppress exponential factors of the correlators whenever convenient, but fully display all prefactors for clarity.
It is suggestive that the prefactors $f_1$ precisely constitute the angular momentum $l$ wave function $\psi_l(z) = z^l   e^{- |z|^2 /4m\ell^2}$ for a charge $e/m$ particle in the LLL. Although we have no formal derivation of 
this, we find below a similar interpretation  in the case of several quasiparticles, 
where their anyonic nature is also manifest.

As pointed out previously, the quasiparticle wave function above is obtained by modifying one of 
the electron operators, rather than inserting a new operator. This is very suggestive of the CF picture 
of a quasiparticle as an excitation of a composite fermion to a higher CF Landau level.
In fact, what  originally led us to construct the operator $P_{\frac{1}{m}}$ was the observation that
the wave function \pref{1qp} is  {\em identical} to the corresponding CF wave function 
(Eq. 5 of ref. \onlinecite{jain3}), which is known to have a good variational energy and the 
correct fractional charge.
In spite of this identity, however, there are two differences between the present
derivation and the CF construction that deserve to be noted: First, the present formalism 
is entirely within the lowest Landau level.
The CF construction of wave functions, on the other hand, involves placing composite
fermions in higher CF Landau levels and subsequently projecting onto the LLL
by replacing all $\bar z$:s by derivatives in the resulting polynomial. 
Technically, of course, when deriving the one-quasiparticle wave function,
the derivatives in (\ref{1qp}) enter in the exact same places as those
due to projection in the CF construction -- but no projection is 
needed in the present formalism\cite{hax}. We return to this point in section \ref{sec:IV},
where we construct the ground states of the Jain sequences at $\nu = n/(2np+1)$.
\footnote{ As straightforward projection tends
to get computationally heavy in numerical calculations with many particles and
a large number of derivatives, slightly different methods of obtaining LLL wave functions
have been employed in most of the CF literature\cite{review}. These, too,
are often referred to as projection. For a single quasiparticle the different prescriptions 
agree, while in the general case they produce very similar but not identical wave functions. 
It is the brute force projection which exactly matches with the CFT construction
for the Jain sequence ground states.}
Second, in spite of the close relation to composite fermions, the operator 
$P_{\frac{1}{m}}(z) $ is not fermionic, as can be seen from the commutation relation 
$P_{\frac{1}{m}}(z) P_{\frac{1}{m}}(w) - e^{i\pi (m-2+1/m) } P_{\frac{1}{m}}(w) P_{\frac{1}{m}}(z) =0$   
or the OPE  $P_{\frac{1}{m}}(z)  P_{\frac{1}{m}}(w) \sim (z-w)^{m-4+ \frac 1 m}\,  
e^{i \frac {2(m-1)} {\sqrt m} \varphi_1(w)} $, that follow from \pref{verrel}. 
The precise connection to composite fermions will be discussed in the section on the 
$\nu = 2/5$ state below. Although the fractional exponent $1/m$ suggests fractional 
statistics, one cannot directly read the statistical angle from the two-point function. 
This issue is discussed in more detail in section \ref{sec:V}. \\

\subsection{Two or more quasiparticles} \label{sec:IIC}
Based on the experience with the single quasiparticle case, we expect the wave function for $M$ quasiparticles to be of the form
\be{nqp1}
\Psi_{Mqp}^{(l)}  (z_i) &=&  {\cal A}\{  f_M(z_1\dots z_M)  \av{ \vtm 1\dots \vtm M \vm {M+1}  \dots  \dots    \vm N     } \} .
\ee
The form of $f_M$ is determined by the condition that the final electronic wave function 
be analytic and antisymmetric, with limiting
behavior $\sim(z_p - z_q)^{m-1+l_{pq}} $, with the relative angular momenta $l_{pq} \geq 1$ and odd.  Because the correlator gives non-analytic factors of the type
$\partial_p \partial_q (z_p - z_q)^{m-2+1/m}$ from all contractions among quasiparticle operators, we choose
$$ f_M (z_1\dots z_M) = g(Z) \prod_{p<q}^M \jas p q {1 + l_{pq}-1/m }
 e^{- \sum_i^M |z_i|^2 /4m\ell^2} \, ,
$$
where $Z = \frac 1 N \sum_{i = 1}^N z_i$ is the center of mass coordinate.
Again, the exponential factors are included to give the correct gaussian factor
$\exp[- \sum_j |z_j|^2 /(4\ell^2)]$ in the $N$-electron wave function. As anticipated 
in the case of one quasiparticle, $f_M$ is just the LLL wave function of
$M$ anyons with fractional charge $e/m$.

To cast  \pref{nqp1} in a form suitable for computation, we will use the following formula, which generalizes  
the expansion by a row used in \pref{1qp} above:
\be{man}
& {\cal A} &\{  \prod_{p<q}^M \jas p q {1-1/m + l_{pq} }    \vtm 1\dots \vtm M \vm {M+1}  \dots  \dots    \vm N  \} \\
 &\sim&  \sum_{ \{i_n\} } (-1)^{\sum_{p=1}^M i_p   } {\cal R} \{ \prod_{p<q}^M  \jas {i_p} {i_q} {1-1/m + l_{pq} }         
   \vtm {i_1}\dots \vtm {i_M}   \vm {\bar i_{M+1}}\dots \vm {\bar i_N}      \}, \nonumber
\ee
where the sum is over all subsets $\{i_1 \dots i_M \} $ of $M$ of the $N$ integers,  and   $\{\bar i_1 \dots \bar i_M \} $ is the conjugate subset of $N-M$ integers. The proof is found in Appendix \ref{app:B1}. 

Using this result, the wave functions for two quasiparticles with total angular momentum $L$ and relative angular momentum $l$ can be written as  
\be{2qp}
\Psi_{2qp}  (z_i) =
&=&\sum_{i<j} (-1)^{i+j}   Z_{ij}^L  \jas i j {1+l-\frac 1 m} e^{ -\frac 1 {4m\ell^2}( |z_i|^2  + |z_j|^2  )}
 \av{ \vtm i \vtm j \prod_{k\ne i,j} \vm k   } ,
\ee
where $Z_{ij} = (z_i + z_j)/2$. 
Evaluating the correlator we obtain the following explicit form for the wave function for two quasiparticles with 
relative angular momentum $l$ and center of mass angular momentum $L$,
\be{2qpe}
\Psi_{2qp}^{l,L}  (z_i) &=& \sum_{i<j} (-1)^{i+j} Z_{ij}^L  \jas i j {1+l-\frac 1 m}  \partial_{z_i} \partial_{z_j} \jas i j {m-2+ \frac 1 m}  \\
&& \prs k  {ij}  \jas k i {m-1}   \prs l  {ij} \jas l j {m-1}  \prjas m n {ij} m,  \nonumber
\ee
where the derivatives act on the whole expression to their right, and
$
\prs k {ij} = \prod_{\stackrel {k=1} {k\ne i,j}}^N
$  and 
$
\pr i j {kl}  = \prod_{\stackrel {i< j} {i,j\ne k,l}}^N  $. 

\noindent
The corresponding wave function in the CF approach is given by\cite{jain3}
\be{jain2qpe}
\tilde\Psi_{2qp}^{l,L}  (z_i) &=& \sum_{i<j} (-1)^{i+j} Z_{ij}^L  \jas i j {l}  \partial_{z_i} \partial_{z_j} \jas i j {m-1} \\
&& \prs k  {ij}  \jas k i {m-1}   \prs l  {ij} \jas l j {m-1}  \prjas m n {ij} m \, .  \nonumber
\ee
The two wave functions differ by terms wherein the derivatives in  \pref{jain2qpe} act on the factor $\jas ij {1-\frac 1 m}$.  It is known \cite{jain3} that the CF wave function in  
\pref{jain2qpe} gives the correct 
fractional charge and statistics of the two-quasiparticle state. 
The first non-trivial test of our construction is therefore to check whether
the CFT wave function  \pref {2qp} shares these good charge and statistics properties.
This is indeed the case, as demonstrated by our numerical simulations, 
which are summarized in section \ref{sec:V} below.
These results show that the two wave functions are essentially identical (for example, their overlap is
99.96\% for 50 particles).
This can be understood from the following heuristic arguments: 
First, since the derivatives in  \pref{jain2qpe} act on a function which is a polynomial of order $N$ in 
both  $z_i$ and $z_j$, this will generate $O(N^2)$ terms. It is unlikely that the few terms picked up by acting on the first factor will be significant.
Secondly, these terms are sub-leading in the coordinate difference $\jas i j {} $ between the quasiparticles, 
and thus unlikely to affect qualitative properties.

\subsection{Quasiparticles and quasiholes}
Wave functions for pairs of quasiparticles and quasiholes can be constructed by
inserting pairs of the corresponding operators into the CFT correlator 
for the Laughlin ground state.
The simplest case is a quasiparticle at the origin together with a quasihole at
position $\eta$, given by
\be{qp-qh}
\Psi_{qp-qh} (z_i, \eta) &=&  {\cal A}\{e^{- |z_1|^2 /4m\ell^2}  \av{ \vtm 1 \vm 2  \dots  \dots    \vm N \, H_{\frac 1 m}(\eta)    } \}\\ 
&=&\sum_i(-1)^{i+1} \, e^{- |z_i|^2 /4m\ell^2}  \av{ \vtm i \prod_{j\ne i} \vm j \, H_{\frac 1 m}(\eta) }  \nonumber \\
&=&\sum_i (-1)^i  \pr j k {i} (z_j-z_k)^m \prod_{j \neq i} (z_j - \eta) \, 
\partial_i  \prod_{l \neq i}(z_l-z_i)^{m-1} \, (z_i - \eta)^{1 - \frac 1 m}, \nonumber
\ee
where the antisymmetrization acts on the electron coordinates $z_i$ only.
More generally, a quasiparticle localized at some position $\eta'$ away from the origin 
may be constructed as a coherent superposition of the angular momentum states given
in  \pref{1qp}. 

For states with equally many quasiparticles and quasiholes, the background charge does not have to be changed from its ground state value. In this sense, wave functions of this type are the natural low energy bulk excitations that do not require any compensating edge charge. On a closed surface, no fractionally charged states are allowed. 

\section{ Composite Fermion states in the Jain series} \label{sec:III}
\subsection{The $\nu = 2/5 $ composite fermion ground state}
\newcommand\tphie{\varphi_2 (z)}
\newcommand\tphi[1]{\varphi_2 (z_{#1})}
In the composite fermion picture, the ground state wave functions at fillings 
$\nu = n/(2np+1)$
are constructed as $n$ filled Landau levels of composite fermions with $2p$ flux quanta
attached. In particular, the $\nu = 2/5 $ state corresponds to filling the lowest
two CF Landau levels. This state may thus be viewed as a ``compact" state of
$N/2$ quasiparticles, \ie the CF:s in the second Landau level
are in the lowest possible total angular momentum state. 

To explore the connection to our CFT construction, we 
generalize the two-quasiparticle wave function \pref{2qpe} of the 
$1/m$ state to the $M$-quasiparticle case, with $M=N/2$, 
and consider a maximum density circular droplet obtained by putting all the
quasiparticle pairs in their lowest allowed relative angular momentum ($\ell =1$), and with zero angular 
momentum for the center of mass ($L=0$). For simplicity we shall also take $m=3$ (and suppress 
 the subscript $m$ on the operators)  since the generalization to arbitrary odd $m$ is 
 obvious. Using \pref{man} and evaluating the correlators, the wave function for $M$ quasiparticles reads
\be{nqp}
\Psi_{Mqp} (z_i) &=& \sum_{i_1<i_2< \dots <i_M} (-1)^{\sum_k^M i_k}  
\prod_{k<l}^M  \jas {i_k} {i_l}  {\frac 5 3}  \partial_{z_{i_1}} \partial_{z_{i_2}} \dots \partial_{z_{i_M}}   
\prod_{k'<l'}^M  \jas {i_k'} {i_l'}   {\frac 4 3} \\
&& \prs {k_1}  {i_2,i_3\dots i_M}  \jas {k_1}  { i_1}  {2}    
\prs {k_2}  {i_1, i_3\dots i_M}  \jas {k_2}  { i_2}  {2}    \dots
\prs {k_n}  {i_1,i_2\dots i_M}  \jas {k_{M}}  { i_M}  {2}    
 \prjas m n  {i_1, i_2\dots i_M}    3  \, .   \nonumber
\ee
Since the anyonic wave function on the first line has the form of a Jastrow factor, it is natural to introduce a second free bosonic field $\tphie$. In fact, by defining
\be{vtidel}
\tilde V(z)  = e^{i\sqrt{\frac 5 3}\tphie}  \partial   e^{i\frac 2 {\sqrt 3} \varphi_1(z)} \, ,
\ee
we find that  \pref{nqp} may be written
in the following compact form
\be{nqpct}
\Psi_{Mqp}  (z_i) &=& {\cal A} \{  \av  { \prod_{i=1}^M \tilde V(z_i)   \prod_{j=M+1}^{N} \vmn j  }    \}
\ee
\ie as a sum of correlators of $M$ $\tilde V$:s and $(N-M)$ $V_1$:s.

Again, this expression differs from the corresponding CF wave function only in the ordering of 
the derivatives and the Jastrow factors in the first line of \pref{nqp}.  
Indeed, as demonstrated in Appendix \ref{app:B}, the CF wave function is obtained simply by moving 
all the derivatives all the way to the left. 
Let us therefore define
\be{vtil}
V_2(z)  = \partial   e^{i\frac 2 {\sqrt 3} \varphi_1(z)}  e^{i\sqrt{\frac 5 3}\tphie}  \, ,
\ee
where the derivative now acts on both the exponentials, 
and consider the case $N=2M$.
We then find that the following sum of correlators of $M$ $V_2$:s and $M$ $V_1$:s:
\be{nqpc2/5}
\Psi_{2/5}^{\rm CF} (z_i) &=& {\cal A} \{  \av  { \prod_{i=1}^M V_2(z_i)   \prod_{j=M+1}^{2M} \vmn j  }    \} \\
&=& \sum_{ \stackrel {i_1<i_2 \dots i_M} {\bar i_1<\bar i_2 \dots \bar i_{M} } } (-1)^{\sum_k^M i_k}    \av{              
V_2 (z_{i_1}) \dots    V_2 (z_{i_M})    \vmn {\bar i_1}   \dots  \vmn {\bar i_{M} }     }  \nonumber
\ee 
exactly reproduces the $(N=2M)$-electron CF wavefunction for $\nu = 2/5$.

The operators $V_2(z_i)$, as opposed to the  $P(z_i)$:s, are
real fermionic operators in that they anticommute among themselves, but 
commute with $V_1(z_i)$:s, just as the
 $P(z_i)$:s. Note that the form of  $V_2$ was determined entirely from the form of  the maximum density 
 $M$-quasiparticle wave function, so  its fermionic nature was not an input. 
 If we want to interpret $V_2$ as a composite {\em electron} operator, it should have the same charge as $V_1$.
 This is ensured if we redefine the charge density operator as 
 \be{modcdo}
J(z) =\frac i {\sqrt 3} \partial \varphi_1 (z)  +   \frac i {\sqrt{15}} \partial \varphi_2 (z). 
\ee
This construction may seem {\em ad hoc} in the sense that we fix the coefficient of $\varphi_2$ by hand 
so as to obtain the correct charge.  However, we shall see below that this choice is
consistent, in that it produces the correct charge for the quasiholes in the $\nu = 2/5$ state. 

Fulfillment of  the charge neutrality condition for the vertex operators $V_2$ requires a background charge, which for the maximum density circular droplet can be assumed to be constant. Furthermore, this density must reproduce the correct  exponential factor for electrons in the LLL. 
The latter is achieved by redefining the expectation value as 
\be{2/5}
\av{ \cdots }_{2/5 } \equiv  \bra 0 \dots e^{-i\sqrt {15}\tilde \rho_3 \int_{\tilde A} d^2z\, \varphi_2(z) }    
 e^{-i\sqrt 3 \rho_3 \int_A d^2z\, \varphi_1(z) }      \ket 0,
\ee
where $\tilde\rho_3= (1/15) \rho_0$, so the total background electron density is 
$(1/3 + 1/15)\rho_0 =( 2/5)\rho_0$. We stress that this value is not an input, but follows from demanding that $V_2$ describe unit charge particles in the LLL, which was what led us to the above form \pref{modcdo} of
the charge density operator. We now show that this state is indeed homogeneous, \ie that the droplets formed by the $N/2=M$ $V_1$:s and the $M$ $V_2$:s have the same area. Charge neutrality gives the following conditions on
the areas $A$ and $\tilde A$ integrated over in \pref{2/5}, 
\be{cneut}
\sqrt 3 M + \frac 2 {\sqrt 3} M &=& \sqrt 3 \rho_3 A \\
\sqrt{\frac 5 3}M  &=& \sqrt {15} \tilde\rho_3 \tilde A \nonumber \, ,
\ee
which implies $A = \tilde A$  and thus homogeneity. 
From the perspective of composite fermions, this correponds to two filled CF Landau levels, since the degeneracy is the same in all Landau levels.  
It would be interesting to redo the above construction on a closed manifold, where we would expect the concept of ``filled CF Landau level" to emerge in a natural way from the condition that the correlators do not vanish. 

Although it is possible to write general many-quasiparticle wave functions similar to the two particle wave function in  \pref{2qp}, it is only the maximum density droplet 
of  \pref{nqp}, and more generally the "compact" CF states\cite{review}, that allow for a simple expression in terms of conformal blocks as in  \pref{nqpct}; for general relative angular momenta one still has to explicitly put in  compensating (anyonic) wave functions by hand.  In this general case, there is also no reason for introducing a constant background charge different from that of the ``parent" $\nu =1/3$,  so there is no natural way to obtain non-zero correlators even if we were to introduce the field $\tphie$. As we see below, this would also be in conflict with the known properties of the charge 1/3 quasiholes.

\subsection{The quasihole operators}
To create quasiholes in the 2/5 state, the operator $H_{\frac{1}{3}}(\eta)$
of  \pref{hole} is no longer appropriate since it does not give holomorphic 
electron wave functions, as is seen from, \eg, $\langle V_2(z) H_{1/3}(\eta) \rangle \sim (z-\eta)^{2/3}$.  Instead, it is necessary to include the second Bose field, $\varphi_2$, and construct
quasihole operators of the form $H_{pq}(\eta) =
e^{i \frac{p}{\sqrt {3}} \varphi_1 (\eta) + i \frac{q}{\sqrt{15}} \varphi_2 (\eta) } $. 
The coefficients $p$ and $q$ are 
determined from the requirements that (i) the wave function of any single 
quasihole be holomorphic, \ie the power of the correlator between any 
quasihole operator and
$V_1(z)$ or $V_2(z)$ be a non-negative integer, 
and (ii) the resulting hole operator {\it not} be expressible as a combination 
(product) of the other quasihole or vertex operators.
These conditions uniquely determine the allowed coefficients $p$ and $q$, and lead to the
following two fundamental quasihole operators for the $\nu = 2/5$ state:
\be{hmod}
H_{01} &=&  
   e^{ i\frac {3}{\sqrt {15} } \varphi_2 (\eta) }        \\
H_{10} &=&   e^{ \frac i {\sqrt 3} \varphi_1 (\eta) -\frac {2i} {\sqrt {15} } \varphi_2 (\eta) }    \nonumber \ .
\ee
Using the charge operator corresponding to the charge density \pref{modcdo} one verifies that 
both these operators create quasiholes with charge 1/5. 
Note that this charge assignment is a prediction of our scheme, rather than an input, 
since the form of the charge operator \pref{modcdo} was determined independently
from demanding $V_2$ to have unit charge. All other allowed vertex operators can
be constructed as products of $H_{01}(\eta)$ and  $H_{10}(\eta)$; the
operators in \pref{hmod} span the charge lattice.

It is an easy exercise to 
construct the explicit electron wave functions obtained by inserting the operators 
\pref{hmod} in the correlator \pref{nqpc2/5}. Not surprisingly, a
direct correspondence with the composite fermion picture is again 
revealed: Inserting the
operator $H_{10}(\eta)$ (with $\eta = 0$ for simplicity) into the $\nu = 2/5$
ground state \pref{nqpc2/5} exactly gives the wave function of a quasihole in the
center of the lowest CF Landau level, while $H_{01}$ gives a quasihole in the
second CF Landau level.
Taking the product of the two quasihole operators, one obtains a charge-2/5 operator
which, in the CF language, reproduces the wave function of a vortex, \ie
(for $\eta = 0$) two quasiholes at the origin, one in each CF-Landau level. 
\cite{review}
Section \ref{sec:V} clarifies the relation between these
quasihole operators and Wen's effective bulk and edge theories for the 
$\nu=2/5$ quantum Hall state.
 
If we would attempt to use the operators $V_1$ and $V_2$ to describe a 1/3 state with a small 
number of quasiparticles (\eg by putting a compensating charge at the edge or at infinity by hand), 
we would be forced to use the operators \pref{hmod} for the quasiholes and thus be led either to a 
wrong charge assignment for the quasiholes or to redefine the charge operator as to make the $V_2$:s carry 
fractional charge. This again stresses that the form of the charge operator as well as the various vertex 
operators is intimately tied to the particular ground state under consideration.

\subsection{The quasiparticle operator}
The quasiparticle operator of the $\nu = 2/5$ state is constructed in the same spirit as $P_{\frac 1 3}$ given in \pref{qpo}, \ie  as a combination of an ``inverse" quasihole operator and one of the electron operators,
combined with an appropriate number of derivatives. Since in the 2/5 state there are two independent
hole operators ($H_{01}$ and $H_{10}$ in \pref{hmod}) and two electron operators ($V_1$ and $V_2$),
it superficially looks as if as there are four quasiparticle candidates. However, it can be shown \cite{HHV}
that three of these are excluded as they do not produce non-zero wave functions, and one
is left with
\be{2/5qp}
P_{2/5}(z) = \partial^2 e^{\frac{2i}{\sqrt 3}\varphi_1(z) +\frac{2i}{\sqrt{15}}\varphi_2(z) }
\ee
which corresponds to combining $H_{01}$ (a quasihole in the second CF Landau level) with
$V_2$ (a composite fermion in the second CF Landau level). Again, the two derivatives are necessary
in order to produce a non-zero wave function
\be{2/5qpwf}
\Psi_{1qp} (z_i) = {\cal A}   \av  { P_{2/5} (z_1) \prod_{i=2}^{M+1} V_2(z_i)   \prod_{j=M+2}^{2M+1} \vmn j  } ,
\ee
and \pref{2/5qpwf} is identical to the corresponding CF wave function. Note that, given the
connection to composite fermions, it is very natural to have two different quasihole
operators but only one quasiparticle operator: There are two filled CF 
LLs in which to create quasiholes, but the only way (except for higher excitations) to create a quasiparticle is to put one composite fermion  in the third CF Landau level.
 
\subsection{The $\nu = 3/7$ state and the Jain series} \label{sec:IIID}
\newcommand\dt[1]  {\tilde{\tilde {#1}}}
As a final explicit example, let us construct the ground state and quasiholes of
the $\nu = 3/7$ state, \ie the third level of the $\nu = s/(2s+1)$ Jain sequence.
The generalization to the full Jain series is given in Appendix \ref{app:B3} .
 
The 3/7 state is obtained from a correlator containing an equal number of $V_1$:s, $V_2$:s 
and the new operator $V_3$:
  \be{vthree}
 V_3 (z) =  P_{2/5}(z)  e^{i\,  \frac 7 { \sqrt{35} } \varphi_3 (z)  }   =  \partial^2  e^{ i [\frac 2 {\sqrt 3} \varphi_1(z)    
 + \frac 2 {\sqrt {15} }  \varphi_2(z) + \frac 7 {\sqrt{35}} \varphi_3 (z) ] }
 \ee
and again, the result is precisely the $\nu = 3/7$ CF wave function
(see  appendix \ref{app:B3}). 
The relevant charge density operator, which ensures unit charge of $V_3$, is given by 
 \be{modcdo2}
J(z) =\frac i {\sqrt 3} \partial \varphi_1 (z)  +   \frac i {\sqrt{15}} \partial \varphi_2 (z)  
+   \frac i {\sqrt{35}} \partial \varphi_3 (z).
\ee
It is easy to check that $V_3 (z) $ is fermionic, but commutes with both $V_1$ and $V_2$, and that
the wave function written in analogy with \pref{nqpc2/5} has filling fraction $\nu = 3/7$.  
In the language of composite fermions, this corresponds to filling up three CF Landau levels.  In analogy with the 2/5 state, one finds three independent charge-1/7 quasihole operators, which exactly correspond to quasiholes in the third, second, and first CF Landau levels, respectively:
\be{h37}
H_{001}(\eta) &=& e^{i\left[ \frac {5}{\sqrt{35}} \varphi_3(\eta)\right] } \nonumber \\
H_{010}(\eta) &=& e^{i\left[\frac {3}{\sqrt{15}}\varphi_2(\eta) - \frac {2}{\sqrt{35}} \varphi_3(\eta) \right]} \\
H_{100}(\eta) &=& e^{i\left[\frac 1{\sqrt 3} \varphi_1(\eta) -\frac 2 {\sqrt {15}} \varphi_2(\eta) - \frac 2{\sqrt{35}} \varphi_3(\eta) \right]} \, . \nonumber
\ee
Operators for excitations with higher charge 
are obtained as products of these; for example, the product
of all three is a charge-3/7 vortex. Again, it is straightforward to
check that the operators \pref{h37} span the charge lattice. In direct generalization
of the $\nu=2/5$ case, the $\nu=3/7$ quasiparticle operator is given by a combination of the 
inverse hole operator in the highest occupied CF Landau level, \ie $H_{001}$,
and $V_3$, with one additional derivative,
 \be{P37}
  P_{3/7}(z)   =  \partial^3  e^{ i [\frac 2 {\sqrt 3} \varphi_1(z)    
 + \frac 2 {\sqrt {15} }  \varphi_2(z) + \frac 2 {\sqrt{35}} \varphi_3 (z) ] } \, .
 \ee
 
The pattern for construction of higher level operators in the $\nu = s/(2s+1)$ series should now be obvious, and in Appendix \ref{app:B3} we give the general expressions for the operators $V_{pn}$ describing the electrons at the $n^{\rm th}$ level in the $n/(2np+1)$ series, as well as the corresponding current density operator. The proof that the CF wave functions for $n$ filled CF Landau levels are reproduced by sums of correlators with an equal number of  $V_{pn}$:s (for fixed $p$) is outlined in Appendix \ref{app:B3}.
The construction of the pertinent quasihole operators should be straightforward, although we have not derived the explicit formulae beyond the ones given above. 

From the general expressions of the operators, it is easy to see that two operators $V_{pn}(z_i)$ and $V_{pn}(z_j)$ at the same level give a factor $\jas i j {2p + 1}$ in the correlation function, while two operators $V_{p,n_1}(z_i)$ and $V_{p,n_2}(z_j)$ at different levels produce a factor $\jas i j {2p}$
(see appendix \ref{app:B3}). This gives an alternative way to calculate the filling fraction, and also demonstrates that the limiting value for $n \rightarrow \infty$ is $\nu  = 1/2p$.

\section{Connection to effective Chern-Simons theories and edge states}\label{sec:IV}

Wen  has developed a general effective theory formalism for
the QH liquids based on representing the currents by two dimensional gauge fields
$a_{I\mu}$  with a Chern-Simons action\cite{wen}, 
\be{cst}
\cL=- {\frac{1}{ 4 \pi}}\La_{II^\prime}
a_{I \mu} \partial_\nu a_{I^\prime \lambda}~\veps^{\mu\nu\lambda} -
{\frac{e}{2 \pi}} A_{ \mu} \partial_\nu t_I a_{I \lambda}\veps^{\mu\nu\lambda} \, ,
\ee
where  the matrix $\La$ and the ``charge vector" $\btt^T=(t_1\dots t_p) $  have integer elements.
The filling fraction is given by  $\nu= \btt^T K^{-1} \btt$.
A generic quasiparticle carries integral charges of the
$a_{I\mu}$ field, and  is thus  labeled by $p$ integers constituting the vector $\bll = (l_1\dots l_p)$.
The electric charge and the statistics of the quasiparticle are given by  
$
q=-e\btt^T \La^{-1}\bll $
and
$
\theta =\pi \bll^T\La^{-1} \bll 
$,
respectively.
This description is not unique; as explained  in reference [\onlinecite{wen}],  an equivalent description is given by
$(K', \btt', \bll') = (WKW^T, W\btt, W\bll)$ where $W$ is an element of $SL(p,Z)$, \ie an integer valued $p\times p$ matrix with unit determinant. 

As an example of the above, the $\nu = 2/5$ state is described by the $K$ matrix and $\btt$ vector,
\be{k1}
K_{2/5} = 
\left(
\begin{array}{ccc}
 3 &   2   \\
  2 & 3    
\end{array}
\right) \ \ \ \ \ \ \ \ \  \btt^T = (1,1) \, .
\ee
This is an example of what Wen refers to as the symmetric basis, where in general $\btt^T = (1,1,\dots ,1)$. By an $SL(2,Z)$ transformation, we can represent the same state in the ``hierarchy basis" (which naturally occurs when constructing states in the Halperin-Haldane hierarchy) characterized by  $\btt^T = (1,0,\dots ,0)$. 
\be{kprime}
K'_{2/5} = WKW^T = 
\left(
\begin{array}{ccc}
 3 & -1     \\
  -1 & 2    
\end{array} \right) \ \ \ \   ;  \ \ \ \
 \btt'^T = \btt^T W^T= (1,0) \, 
 \ \ \ \  ;   \ \ \ \ \
W = 
\left(
\begin{array}{ccc}
 1 & 0     \\
  -1 & 1   
\end{array} \right)   \, ,
\ee

Starting from the Chern-Simons theory \pref{cst} defined on a finite two dimensional domain, one can derive a dynamical theory for the edge excitations. The details can be found in [\onlinecite{wen}] and references therein, and the resulting theory is
\be{edge}
S_{ed} =  \frac 1 {4\pi} \int dt dx\,[ K_{IJ}\partial_t \phi_I \partial_x \phi_J        - V_{IJ} \partial_x\phi_I \partial_x \phi_J  +  2eA_\mu\epsilon^{\mu\nu} \partial_\nu t_I \phi_I ],
\ee
where $K$ and $\btt$, as well as the quasiparticle vector $\bll$, are the same as in the effective bulk theory \pref{cst}.  This is a multicomponent chiral boson theory with the current operator given by
\be{cscurr}
J^\mu = - \frac {\delta S} {\delta A_\mu} = -\frac e {2\pi} \epsilon^{\mu\nu} t_I  \partial_\nu\phi_I \, .
\ee
 The quasiparticle operators (including the electron operator) take the generic form
\be{qpop}
 \Psi \sim e^{i\sum_q l_q \phi_q } \, ,
 \ee
 familiar from abelian bosonization of one-dimensional fermion systems. The 
numbers $V_{IJ}$ are the non-univeral edge velocities, which depend on the details of the confining potential.

In their original paper on the connection between QH liquids and conformal field theories, Moore and Read made two basic claims. The first, which we already have discussed, is that the electronic wave functions can be expressed as conformal blocks of certain CFT:s. The second is that this very same CFT {\em is} the one dimensional theory describing the dynamical edge excitations. This last claim should not be taken literally since 
 it is known that the edge dynamics is non-universal. Not only the edge velocities, but also the character, and even  the number of edge modes can  depend on details of the edge potential. Examples are the polarization edge modes related to edge spin texture\cite{karlhede}  and the counter-propagating modes resulting from edge reconstruction as first discussed by Shamon and Wen\cite{shamonwen}. Thus we can only hope that the CFT will provide a ``minimal" edge theory consistent with the topological properties of the bulk, \ie that it supports excitations with the same charges. 
In spite of these limitations, the Moore-Read conjecture about the edge theory has been very fruitful, especially in the search for effective field theories for the non-abelian Pfaffian state\cite{frad}. 

We shall now demonstrate the connection between the CFT construction of the Jain states and Wen's 
$K$-matrix formulation by explicitly working out the case of $\nu  = 2/5$. Led by the Moore-Read
conjecture, we will start from our CFT bulk theory, read off the $K$-matrix and the charge vector, and show
that in the basis where \pref{hmod} are the fundamental quasihole operators, one exactly recovers Wen's $K$-matrix
and $\btt$-vector in the symmetric basis. This is consistent with Read's earlier result\cite{read1990} that the
symmetric basis naturally describes the Jain states. Alternatively, we may choose a basis consisting of either
of the charge 1/5 quasiholes in \pref{hmod}, along with the charge 2/5 vortex (\ie the product of the two 1/5-hole
operators); as we shall see, this instead corresponds to the hierarchical basis.

The conformal field theory contains the two uncoupled bosonic fields $\varphi_1$ and $\varphi_2$, compacitfied on radii $R^2 = 3$ and 15, respectively. The corresponding action,  $S = \int d^2x  \, {\cal L}_{\mathrm cft} $ for the full scalar fields $\phi_i (x,t) = \varphi_i (z) + \bar\varphi_i (\bar z)$  is   obtained from the Lagrangian, 
\be{k2}
\cL_{\mathrm cft} = \frac 1 {8\pi} 
(\bar\phi_1, \bar\phi_2)
\left(
\begin{array}{cc}
 1 & 0     \\
  0 & 1   
\end{array} \right)   \partial_\mu \partial^\mu
\left(
\begin{array}{c}
\phi_1   \\
\phi_2  
\end{array} \right)   
+      \frac e {2\pi} \tilde  t_I   A^\mu \,  \epsilon_{\mu\nu} \partial_\nu \phi_I 
 \equiv \frac 1 {4\pi}  K_{IJ} \phi_I  \partial_\mu \partial^\mu \phi_J  
 - A^\mu J_\mu\, .
\ee
where the information about the compactification radii is contained in the  charge vector  $\tilde \btt^T = (1/\sqrt 3, 1/\sqrt{15})$. 
The Lagrangian \pref{k2} contains both right and left moving fields, but these decouple, and it is known that the dynamics of a single chiral component, such as $\varphi_i(z)$,  is described by the first order Lagrangian  \pref{edge} with the same $K$ matrix and ${\bf t}$ vector\cite{fadjack}.    
In order to directly compare with Wen's formalism, we rescale the Bose fields such as to obtain an
integer charge vector, $\btt^T = (1,1)$: $(\varphi_1', \varphi_2') \equiv ( \varphi_1/\sqrt 3 ,  \varphi_2/\sqrt {15}) $.
Naively, the corresponding $K$ matrix would then be $diag(3, 15)$.
It is however important to remember that a CFT  is not defined only by the Lagrangian of the fields $\varphi_i$, which gives the operator product expansions, or fusion rules, of the primary fields (\ie the vertex operators), but also by primary field content, \ie the allowed vertex operators. In the case of  the $\nu = 2/5$ state these allowed fields define a  charge lattice with the basis vectors given by the quasihole operators \pref{hmod}. Thus, we will change to a basis $(\chi_1, \chi_2)$ where the fundamental quasihole
operators spanning the charge lattice are given by $H_i = e^{i\chi_i}$.
As can be seen from \pref{hmod}, this is achieved by the field redefinition,
\be{redef}
\chi_1 &=&  \frac 3  {\sqrt  {15} } \, \varphi_2 = 3 \varphi_2'   \\
\chi_2 &= & \frac 1 { \sqrt 3}\,   \varphi_1  -   \frac 2  {\sqrt  {15}}  \,  \varphi_2  =  \varphi_1'  - 2 \varphi_2'   \nonumber \, .
\ee
Inverting this transformation and inserting into \pref{k2}, it is now easy to verify that the resulting $K$ matrix and $\btt$ vector 
are precisely the $K$ and $\btt$ in \pref{k1}.  Alternatively (and equivalently),
if we start from a basis of one of the 1/5 quasihole operators, say $H_{10}$,
together with the charge 2/5 ``vortex" $H_{11} \equiv H_{10} H_{01}$, corresponding to the change of basis
\be{redef2}
\chi_1 &=&   \varphi_1' +  \varphi_2'   \\
\chi_2 &= & \varphi_1'  - 2 \varphi_2'   \nonumber \, ,
\ee
we find that the corresponding $K$-matrix and $\btt$-vector are the ones given in \pref{kprime}, \ie the hierarchical
basis. This equivalence, at the effective Chern-Simons theory level, of the Jain states and the hierarchy scheme,
has been previously pointed out by several authors \cite{read1990, blokwen}. These authors arrive at this result by a  
general argument, based on similarity between the Jain states and filled Landau levels, that ignores the projection 
on the lowest Landau level.
It is reassuring that the above demonstration, based on explicitly  holomorphic wave functions, leads to the same result.

This construction straightforwardly carries over to the other fractions in the Jain sequence;
for example, in the case of $\nu = 3/7 $, one may pick the three charge-1/7 quasihole operators
of \pref{h37} as basis of the charge lattice, corresponding to the field redefinition
\be{redef37}
\chi_1 &=& \frac 5 {\sqrt{35}} \varphi_3  \nonumber \\
\chi_2&=&  \frac 3 {\sqrt{15}} \varphi_2 -  \frac 2 {\sqrt{35}} \varphi_3   \\
\chi_3 &=&  \frac 1 {\sqrt 3} \varphi_1 -\frac{2}{\sqrt{15}} \varphi_2 -  \frac 2 {\sqrt{35}} \varphi_3  \nonumber \, .
\ee
Again, this brings us to the symmetric basis, with $\btt^T = (1,1,1)$ and the $K$-matrix given by $K_{ij}=2 + \delta_{ij}$.
Alternatively, we may construct a basis consisting of quasihole operators with charge 1/7, 2/7 and 3/7, respectively,
by appropriate combinations of the charge-1/7 quasiholes in \pref{h37}. As before, this corresponds to the hierarchical
basis, with $\btt^T = (1,0,0)$ and the same $K$-matrix as that given by Wen\cite{wen},
\be{k37}
K_{3/7}^{h} = 
\left(
\begin{array}{ccc}
 3 &   -1 & 0   \\
  -1 & 2 & -1 \\
  0 & -1 & 2    
\end{array}
\right) .
\ee

\section{Localized quasiparticles and fractional charge and statistics} \label{sec:V}
The present formulation already gives a strong hint for fractional charge and fractional 
statistics of the CF quasiparticles: We have seen from \pref{elnr} that  the operator $P_{\frac{1}{m}}(z)$ corresponds to a localized charge at $z$, and the presence of the factor $(z_i - z_j)^{\frac 1 m}$ is suggestive of fractional statistics with angle $\frac \pi m$. This is not a proof, however.  The usual argument for fractional charge and statistics
proceeds via the Berry phases produced by 
adiabatic braidings of localized quasiparticles. In this section we construct the wave functions for localized  states of one and two quasiparticles, and use these to calculate the Berry phases relevant for charge and statistics within what we call a ``random phase assumption". 

A localized quasiparticle state is constructed as a coherent superposition of a the angular momentum states given in \pref{1qp} and \pref{2qpe}. For a single  quasiparticle at location $\eb$ we have (putting $\ell = 1$) 
 \be{oneloc}
\Psi_{1qp} (\eta ,\eb ; z_i) &=&\tilde {\cal N}_1(\eb \eta) e^{-\frac 1 {4m} |\eb |^2 } \sum_{l = 0}^\infty  
   \frac {  {\bar \eta} ^{l}  }   {(2m)^ll!} \Psi_{1qp}^{l}  (z_i)  \\ 
   &=& \tilde{\cal N}_1(\eb\eta) \sum_i (-1)^i e^{-\frac 1 {4m} (|z_i|^2 +  |\eb |^2   - 2\eb z_i )     } 
   \av{ P(z_i) \prs j i V(z_j) } 
    \nonumber \, . 
\ee
Notice that the normalization constant $ {\cal N}_1(\eb \eta) =   \tilde{\cal N}_1(\eb \eta)  e^{-\frac 1 {4m} |\eb |^2 }    $ only depends on the combination $\eb\eta$. Likewise, we construct the wave function for  two quasiparticles at positions $\eb_\pm = \beb \pm \eb/2$ as 
\be{loc}
\Psi_{2qp} (N,\beb, \eta, \eb ; z_i) 
= \tilde{\cal N}_2(N,\beb, \eta, \eb)  e^{-\frac 1 {8m} |\eb |^2   -\frac 1 {2m} |\beb |^2} \sum_{l = 1,3,...} \sum_{L=0,1,...}  
\frac { (\frac 1 m \beb )   ^L} {L!} \frac { (\frac 1 {4m}  {\overline \eta})^{l-1}  }   {l!} \Psi_{2qp}^{l,L}  (z_i) \\
=   \tilde{\cal N}_2
 \frac {4m} \eb  \sum_{i<j}    (-1)^{i+j}    e^{-\frac 1 {8m} ( |\eb |^2  + |z_{ij}|^2 ) } \sinh{ \frac { \eb \zij}  {4m}}
e^{- \frac 1 {2m} ( |\beb |^2 + |\bzij |^2 - 2 \beb Z)   } 
 \zij^{1-\frac 1 m} \, \av{ \vtm i \vtm j \prs k {ij} \vm k   } \, ,  \nonumber
\ee
where $z_{ij} = z_i -z_j$. For $\eb = 0$ and $\beb = \eb = 0$, respectively, these expressions reduce  to $\Psi_{1qp}^{0}$  and 
$\Psi_{2qp}^{0,0}$, the wave functions with minimum angular momentum. 
The explicit wave functions obtained by evaluating the correlators in \pref{oneloc} and \pref{loc}, are very similar, but not identical, to the corresponding CF wave functions. One source of difference is the slight deviation between the 
angular momentum eigenstates given by \pref{2qpe} and \pref{jain2qpe}, pointed out in section \ref{sec:IIC}, and shown to be numerically insignificant in section \ref{sec:VI}. The other 
source of difference can be seen already for the one quasiparticle state. 
The CF wave function reads 
\be{cf1qp}
\Psi_{CF}=\sum_i(-1)^i\,e^{-\frac{|\eta|^2}{4m}+\frac{\bar{\eta}z_i}{2m}}
\prod_{j<k}^{(i)}(z_j-z_k)^3\prod_n^{(i)}(z_i-z_n)^2
\left[\left(\frac{1}{m}-1\right)\bar{\eta}+\sum_n^{(i)}\frac{4}{z_i-z_n}\right]
\exp\left(-\frac{1}{4}\sum_k|z_k|^2\right), \label{CF1qpWF}
\ee
but the term proportional to $(1/m - 1)\eb$ is missing in the corresponding CFT wave function. This difference, however, is a finite size effect. 
This term contributes to the wave function only when the exponential factor $\exp({\eb z_i/2m})$ is expanded to the $N^{th}$ power, and thus amounts to a 
(nonuniversal) boundary term. 

Before proceeding to the calculation of the charge and statistics of the quasiparticles, it is helpful to recall, as a background, the corresponding calculation for the quasiholes of $\nu=1/m$. Consider the normalized wave function for one quasihole, given by \pref{lqh}: 
\be{1lqh}
\Psi(\eta, \eb; z_i) &=&  {\cal N}'e^{-\frac 1 {4m\ell^2}   |\eta|^2    }   \Psi_L(\eta; z_i) 
\ee
where
$
\Psi_L(\eta;  z_i) =  \prod_i (z_i - \eta_1)   \prod_{k<l} \jas k l m e^{-\frac 1 {4\ell^2}  \sum_i |z_i|^2     }  
$
is the Laughlin unnormalized wave function for a single quasihole.  The plasma 
analogy shows that ${\cal N}'$ is independent of $\eta$; the normalization integral of \pref{1lqh} is the partition function of a Coloumb plasma with a charged impurity, which is independent of the position of the impurity as long as it is farther than a screening length ($\ell$) from the edge.
The Berry phase associated with a circular loop 
$\eta = Re^{i\theta}; \ \theta\in\{0, 2\pi\}$, is given by (with $\hbar=c=1$)
\be{berr}
\gamma_B& =& \int_0^{2\pi} d\theta\,   \bra {\Psi(\eta,\eb) } i\partial_\theta \ket {\Psi(\eta , \eb)  } \\
&=& \int_0^{2\pi} d\theta\,  \bra {\Psi(\eta,\eb) }( \eb\partial_{\eb} - \eta\partial_\eta    ) \ket {\Psi(\eb , \eta)  } \nonumber \\
&=&  -A\frac e m B,   \nonumber
\ee
where $A$ is the area enclosed by the loop, and the last line employs integration by parts and the observation that the only $\bar\eta$ dependence in the wave 
function is from the gaussian factor.  The result, as expected, is  
the Aharanov-Bohm phase for a particle of charge $e/m$. 

Turning to two quasiholes and again using  \pref{lqh} we have 
\be{2lqh}
\Psi(\eb_a,\eta_a; z_i) = {\cal N}''  (\eta_1 - \eta_2)^{\frac 1 m} e^{-\frac 1 {4m\ell^2}  ( |\eta_1|^2  +   |\eta_2|^2  )   }  
\Psi_L(\eta_a; z_i)
\ee
where
$
\Psi_L(\eta_a; z_i) = 
\prod_i (z_i - \eta_1)  \prod_j (z_j - \eta_2) \prod_{k<l} \jas k l m e^{-\frac 1 {4\ell^2}  \sum_i |z_i|^2     }
$ is the unnormalized Laughlin wave function for two quasiholes.
Naively we would read the fractional statistics parameter as $\theta = \pi/m$ directly form the factor $ (\eta_1 - \eta_2)^{\frac 1 m}$ in \pref{2lqh} but that gives the correct result  only if there is no extra Berry phase (other than the usual Aharonov-Bohm phase) associated with the exchange path (here integrating $\theta$ from $0$ to $\pi$)\cite{nw}. 
The absence of additional phases can be confirmed by an explicit calculation 
using the plasma analogy for a system with two impurities  separated farther than the screening distance, $\ell_0$. 
Taking $\eta_1 = -\eta_2 = \eta $, we get $\gamma_B^{ex} = -\frac e m B  \pi R^2$, which is nothing but the AB-phase expected from the exchange of two charge $\frac  e m$ particles through a circular path with radius $R$.  
That allows us to read the exchange statistics phase 
directly from the factor $ (\eta_1 - \eta_2)^{\frac 1 m}$ in \pref{2lqh}. Wilczek and Nayak\cite{nw} suggest that it 
is no coincidence that  \pref{lqh} yields a wave function with no Berry contribution to the statistics angle;  they make a conjecture, supported by arguments, that QH wave functions given directly as correlators, or conformal blocks, of a CFT have vanishing  Berry phase (forgetting the Aharanov-Bohm contribution) so the exchange statistics can be obtained from the so-called monodromy, which in this simple case  is just the phase $e^{i2\pi/m}$ produced by the factor  $ (\eta_1 - \eta_2)^{\frac 1 m}$ when the quasiholes braid around each other along closed paths defined via an analytic continuation of the original wave function. 
In the more general case of non-Abelian fractional statistics, several conformal blocks correspond to the same configuration of quasihole coordinates, and the monodromies are now matrices that encode how these conformal blocks transform into each other under braidings of the coordinates. 

Rather than using \pref{berr}, we follow Kj{\o}nsberg and Leinaas who 
showed that for a (normalized) wave function of the form, 
$\Psi (\eta, \eb; z_i) = {\cal N}(\eb\eta) \Psi_h(\eta,\eb; z_i)$, where ${\cal N}$ is a real function only of $\eb\eta$, the Berry phase is given by \cite{kjons1}
\be{berry2}
\gamma_B  = \int_0^{\Theta} d\theta\,   \bra {\Psi(\eta,\eb) } i\partial_\theta \ket {\Psi(\eta)  } = \Theta R^2 
\frac d {dR^2} \ln {\cal N}(\eb\eta)^2  \, .
\ee
The Laughlin wave function for a single quasihole at $\eta$ or two 
quasiholes at $\pm \eta$ has this form.  
The upper limit is taken to be $\Theta=2\pi$ for a single quasihole, and $\Theta=\pi$ for an exchange of two quasiholes. 
From \pref{1lqh} and \pref{2lqh}, the normalization constants are given by (up to an $\eta$ independent factor) 
\be{qhnorm}
 {\cal N}_1(\eb\eta)   &=&    e^{-\frac 1 {4m\ell^2}   |\eta|^2    }  \\ 
 {\cal N}_2(\eb_i \eta_i) &=&         |\eta_1 - \eta_2|^{\frac 1 m} e^{-\frac 1 {4m\ell^2}  ( |\eta_1|^2  +   |\eta_2|^2  )   },  \nonumber
\ee
so the formula \pref{berry2} can be applied. (We have assumed sufficiently far separated 
quasiholes.) For a single quasihole loop of radius $R$,  \pref{berry2} gives the Berry phase $-2\pi A B (e/m)$.  For two quasiholes at $\pm \eta$, the Berry phase is $(\pi/m)-A B (e/m)$ (with $\Theta=\pi$).  The difference, $\pi/m$, gives the contribution from fractional statistics. 
In this case, there is no monodromy, 
and the the full statistical phase appears as a Berry phase. \footnote{
The quantity ${\cal N}(\eb\eta)$ can be interpreted as the (diagonal element) of the quasihole density matrix,
$
{\cal N} (\eta, \eb)^{2} \equiv \rho (\eta, \eb)
 = \psi(\eta) \psi(\eb)^\star 
$, 
where the factor $\psi(\eta_1,\eta_2)$ is the quasihole wave function. }

We now turn to the case of quasiparticles, where we need to calculate the relevant normalization constants from the wave functions \pref{oneloc} and \pref{loc}. 
Because the sum in \pref{loc} extends only over even powers of $\eb$, the holomorphic part is single valued under $\eb \rightarrow -\eb$,  
implying an absence of monodromy contribution to the statistical angle, 
and thus both the charge and statistics can be extracted directly from 
$ {\cal N}_1 $ and  $ {\cal N}_2 $, provided that they can be chosen  
as real functions of $\eb \eta$ only. 
For quasiparticles, the normalized wave function has the form ${\cal N}(\bar\eta\eta) \Psi(\bar\eta)$, and the Berry phase is given by
\be{berry3}
\gamma_B  =-\Theta R^2 
\frac d {dR^2} \ln {\cal N}(\eb\eta)^2  \, .
\ee
Unfortunately, the calculation of the normalization factors is more difficult than in the case of  the quasiholes, because the electronic wave functions \pref{oneloc} and \pref{loc} involve  sums over correlators; \eg for two quasiparticles the relevant integral is  $\sim \sum_{ijkl}  (-1)^{i+j+k+l} \av{ \vtm i \vtm j \prs p {ij} \vm p }  \av{ \bvtm k \bvtm l \prs q{lm} \bvm j } $. If, however, we keep only the diagonal terms in the sums,
which amounts to a kind of random phase assumption discussed in the next section, 
then the normalization factors can be calculated as shown in Appendix \ref{app:C}. 
The calculation of the normalization constants, outlined in Appendix \ref{app:C}, gives the result:
\be{berres}
 {\cal N}_1^2   &=&      \tilde{\cal N}_1^2  \,       e^{-\frac 1 {2m\ell^2} |\eta|^2}   
 \sim (R^2)^{-1}  e^{-\frac 1 {2m\ell^2} R^2}         \ \ \ \ \  ; \ \ \ \ \   \eta = R e^{i\theta}   \\
\label{berres2}
  {\cal N}_2^2      &=&      \tilde {\cal N}_2^2   \,         e^{-\frac 1 {2m\ell^2} |\eta|^2}   
  \sim   (R^2)^{\frac 1 m -2} e^{-\frac 1 {4m \ell^2} R^2}    \ \ \ \ \  ; \ \ \ \ \   \eta = \frac R 2 e^{i\theta}   
  \ee
Using \pref{berry3} we can extract the fractional charge from the Berry phase corresponding to a single quasiparticle,
\be{fracch}
\gamma_B = \frac e m B \pi(R^2 + 2m \ell^2) = \frac e m BA + 2\pi \, ,
\ee
so the leading term is precisely the expected AB phase for a charge $e/m$ object. The $O(\ell^2)$ term is a quantum correction to the ``classical" $R^2$. Such corrections have been discovered earlier both in the context of CF quasiparticles\cite{jain2} and noncommutative matrix models\cite{poly}. 
The statistical angle is obtained by subtracting this result from the Berry phase extracted from the exchange path.   Remembering that the effective area  enclosed by the exchange path is $\pi R^2/4$, we get
\be{stat}
\theta = \gamma_B^{exc} -  \frac 1 4 \gamma_B = - \frac \pi m \, .
\ee
This reproduces the result obtained from a direct numerical evaluation of the Berry phases from the composite fermion wave functions \pref{oneloc} and \pref{loc}\cite{kjons,jain2,kjons2}.  
The statistics of the quasiholes and quasiparticles obtained above differ in sign, contrary to  
the expectation from general considerations based on effective Chern-Simons or the CF theory, and we want to comment upon this.
Both normalization constants in \pref{berres} are of the form ${\cal N} \sim (R^2)^a e^{-bR^2}$, and from the derivation in the Appendix 
one sees that the constant $a$ is unambiguously determined, and the same holds for the fractional part of  $b$, while the integer part of $b$ could in principle be shifted  by using a different prescription for the ordering of the derivatives. However, this does not necessarily mean that the frational part of $\theta$ is well determined since it is sensitive to a cancellation of a large number of terms proportional to $R^2$ in the Berry phases. The above result is based on the assumption that we have correctly identified $|\eta|$ as the distance between the two quasiparticles.
This question is discussed in some detail in 
in Ref. \onlinecite{jain2} where it is shown that the distance between two quasiparticles is slightly different from $|\eta |$; correcting for the distance produces the same statistics as for quasiholes (modulo an integer).  Because our wave function is closely related to the CF wave function, similar considerations should apply here as well.

\section{Numerical tests}  \label{sec:VI}

This section concerns quantitative tests of the CFT quasiparticle wave 
functions. The numerical tests consist of two parts. In the first part, we compare, at
filling factor $\nu=1/3$, the two-quasiparticle wave functions \pref{2qpe} obtained from the conformal field theory  with the standard wave functions from the CF theory by calculating their Coulomb interaction energies and overlap.  We will find 
that the two are practically identical.  In the second part, the random phase
approximation used in section \ref{sec:V} is examined. 
\subsection{Two-quasiparticle wave function}

The $N$-composite fermion wave function for two quasiparticles 
at filling factor $\nu=1/3$ is constructed by compactly filling the lowest CF-LL by $N-2$ 
composite fermions, and placing the remaining two composite fermions in the second  
CF-LL.  We consider below the state in which the two ``excited" composite fermions 
are in angular momentum orbitals, and occupy the smallest angular momenta.  
The wave function for this state is written as
\be{cfdet}
\Psi_{\rm CF} = {\cal P}_{\rm LLL}\left|
\begin{array}{cccc}
\eta_{1,-1}(z_1)   &  \eta_{1,-1}(z_2)    & \ldots  & \eta_{1,-1}(z_N) \\
\eta_{1,0}(z_1)     &  \eta_{1,0}(z_2)     & \ldots  & \eta_{1,0}(z_N) \\
\eta_{0,0}(z_1)     &  \eta_{0,0}(z_2)     & \ldots  & \eta_{0,0}(z_N) \\
\eta_{0,1}(z_1)     &  \eta_{0,1}(z_2)     & \ldots  & \eta_{0,1}(z_N) \\
\vdots                  &  \vdots                  &             & \vdots \\
\eta_{0,N-3}(z_1) & \eta_{0,N-3}(z_2)  & \ldots  & \eta_{0,N-3}(z_N)
\end{array}
\right|    \times\prod_{i<j} (z_i-z_j)^2\exp\left(-\frac{1}{4}\sum_k|z_k|^2\right),
\ee
where ${\cal P}_{\rm LLL}$ denotes projection into the lowest Laudau level (LLL), and 
$\eta_{n,m}(z)$ is the single particle eigenstate in symmetric gauge: 
\be{norm}
\eta_{n,m}(z,\bar{z})=N_{n,m}\,e^{-|z|^2/4}\sum_{k=k_0}^n(-1)^k
\left(\begin{array}{c} n+m \\ n-k \end{array}\right) \frac{1}{2^k k!} \bar{z}^k z^{k+m},
\ee
in which $n=0,1,2,\ldots$ labels the Landau level index,  
$m=-n,-n+1,-n+2,\ldots$ is the angular momentum quantum number, 
$k_0=\mbox{max}(0,-m)$, and 
\be{}
N_{n,m}=\sqrt{\frac{n!}{2\pi 2^m(n+m)!}}
\ee
is the normalization constant.
As usual, the magnetic length has been set to unity.
The wave function can be shown to be equal to 
\be{}
\Psi_{\rm CF} ={\cal P}_{\rm LLL}\left|
\begin{array}{cccc}
\bar{z}_1 z_1   &  \bar{z}_2 z_2    & \ldots  & \bar{z}_N z_N \\
\bar{z}_1     & \bar{z}_2  & \ldots  & \bar{z}_N \\
1     & 1     & \ldots  & 1 \\
z_1     &  z_2     & \ldots  & z_N \\
\vdots                  &  \vdots                  &             & \vdots \\
z_1^{N_3} & z_2^{N_3} & \ldots  & z_N^{N_3}
\end{array}
\right|    \times\prod_{i<j} (z_i-z_j)^2 \exp\left(-\frac{1}{4}\sum_k|z_k|^2\right).
\ee
Following the standard procedure, the projection procedure is accomplished by expanding the determinant, 
moving all $\bar{z}$'s to the left, and replacing $\bar{z}$ by $2\partial/\partial z$ (with 
the convention that the derivatives do not act on the Gaussian part). 
A technique developed in Ref. [\onlinecite{Kamilla1997}] has made it possible to perform 
the projection in a more convenient manner, which is the one we use below.  (This 
method gives projected states very close to those obtained by ``brute force" projection.)
The projected wave function is written as 
\ba
\Psi_{\rm CF} &=&\sum_{i<j}(-1)^{i+j}(z_i-z_j)^3\prod_k^{(ij)}(z_i-z_k)^2\prod_l^{(ij)}(z_j-z_l)^2
\prod_{m<n}^{(ij)}(z_m-z_n)^3\nonumber\\
&&\times\left\{\frac{-2}{(z_i-z_j)^2}+\sum_{k,l}^{(ij)}\frac{4}{(z_i-z_k)(z_j-z_l)}+
\sum_k^{(ij)}\frac{4}{(z_k-z_i)(z_k-z_j)}\right\}
\exp\left(-\frac{1}{4}\sum_k|z_k|^2\right).  \label{CF2qpWF}
\ea
The symbol $(ij)$ denotes that indices $i$ and $j$ are excluded in the summation or
the product. The total angular momentum of  \pref{CF2qpWF}  is
\be{}
L=\frac3 2 N^2 - \frac 7 2 N + 2. \label{TotalL}
\ee
Numerical simulations for the wave function in \pref{CF2qpWF} have 
shown that it produces 
better variational energy than the two quasiparticles wave function obtained by 
generalizing Laughlin's single quasiparticle state\cite{jain3}.

The CFT wave function for two quasiparticles located at the origin is given by \pref{2qp}.
To make contact with the above CF wave function, we set the center of mass angular momentum to zero and put the 
two quasiparticles in the smallest relative angular momentum channel $l=1$; 
that produces a wave function that has the total angular momentum given in \pref{TotalL} .  For these parameters,  the CFT ansatz for the
two quasiparticles wave function $\Psi_{\rm CFT}$ at $\nu=1/3$ reduces to 
\be{CFT2qpWF}
\Psi_{\rm CFT} &=& \sum_{i<j}(-1)^{i+j}(z_i-z_j)^3\prod_k^{(ij)}(z_k-z_i)^2\prod_l^{(ij)}(z_l-z_j)^2\prod_{m<n}^{(ij)}(z_m-z_n)^3\\\nonumber
&&\times\left\{
-\frac{4}{9}\frac{1}{(z_i-z_j)^2}+\frac{8}{3}\sum_k^{(ij)}\frac{1}{(z_k-z_i)(z_k-z_j)}+\sum_{k,l}^{(i,j)}\frac{4}{(z_k-z_i)(z_l-z_j)}\right\}\exp\left(-\frac{1}{4}\sum_k|z_k|^2\right).
\ea

\begin{figure}
\includegraphics[scale=0.57,angle=-90]{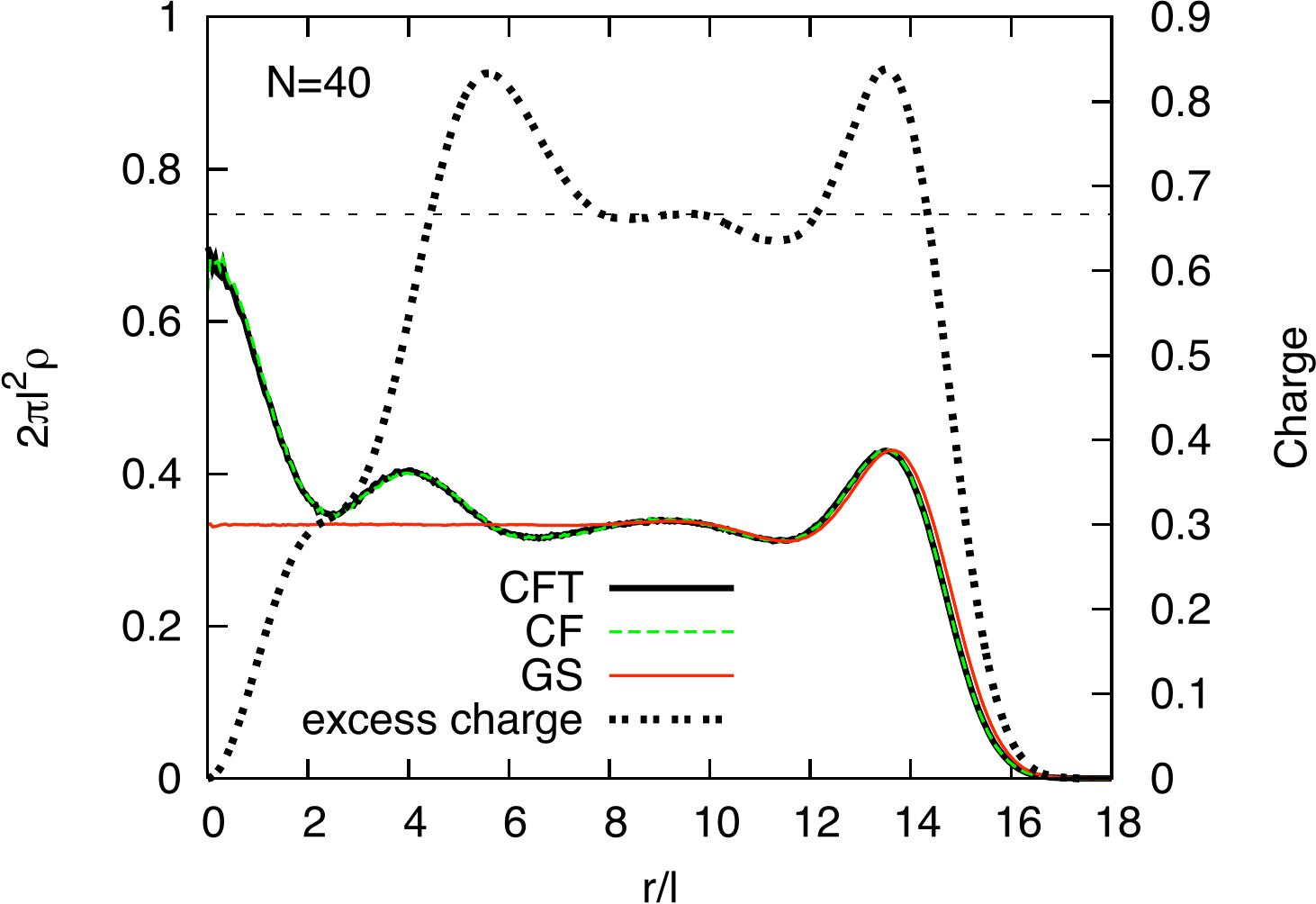}
\includegraphics[scale=0.57,angle=-90]{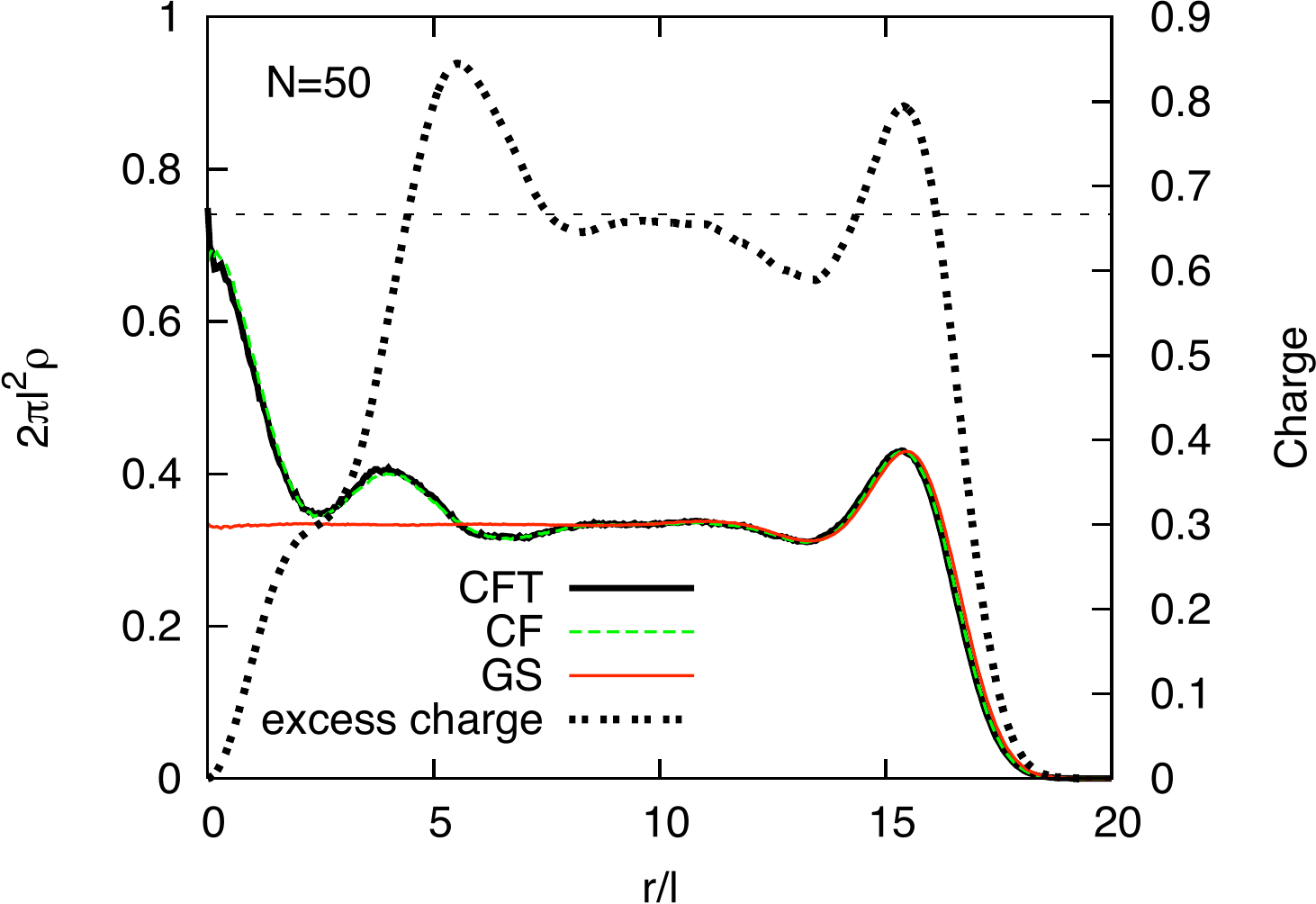}
\caption{Density profiles of $\Psi_{\rm CFT}$ and $\Psi_{\rm CF}$ for $N=40$ 
(left panel), 
and 50 (right panel) particles. ``GS'' denotes the $\nu=1/3$ background obtained from  
Laughlin's wave function. The horizontal dashed line indicates 2/3 units of charge. 
In both cases, the excess integrated charge for $\Psi_{\rm CFT}$, denoted by thick 
dashed line, shows the correct 
quantized value. The statistical undertainty in Monte Carlo is smaller than the widths
of the lines. 
}
\label{DensityProfiles}
\end{figure}

To compare the two-quasiparticle wave functions in Eqs.~(\ref{CFT2qpWF}) 
and (\ref{CF2qpWF}), we compare their Coulomb interaction energies and also calculate 
their overlaps.  The Coulomb energy (in units of $e^2/\epsilon\ell_0$) is defined as
\be{coul}
E= \frac{\left\langle\Psi\left|\frac 1 2\ds\sum_{i\neq j}\ds\frac{1}{|\bfr_i-\bfr_j|}\right|\Psi\right\rangle}{\langle\Psi|\Psi\rangle},
\ee
We do not include here electron-background and 
background-background contributions; these are not necessary for the present purpose, as they are identical for the two wave functions. 
The overlap is defined as
\be{over}
{\cal O}=\frac{|\langle\Psi_{\rm CFT}|\Psi_{\rm CF}\rangle|}{\sqrt{\langle\Psi_{\rm CFT}|\Psi_{\rm CFT}\rangle\langle\Psi_{\rm CF}|\Psi_{\rm CF}\rangle}}.
\ee
Both quantities are evaluated using Metropolis Monte Carlo integration. A single data
point is obtained by averaging 100 independent Monte Carlo runs, with $\sim 1.2\times 10^5$ iterations in each run. For $N=50$ electrons, the total computational time is
approximately 100 hours on a single node of a Beowulf-type PC cluster consisting of dual 
3.06 GHz Intel Xeon Processors. The results are summarized in Table \ref{EnergyAndOverlap}.  The excellent agreement demonstrates that 
the two two-quasiparticle wave functions are essentially identical.
Figure \ref{DensityProfiles} depicts the density profiles of the two 
wave functions for $N=40$ and 50 electrons. The excess 
integrated charge (\ie the charge measured relative to 
the $\nu=1/3$ background) for the CFT wave function 
is also shown. The excess charge has a well-defined plateau at 2/3 before the edge 
distortion.  (The creation of two quasiparticles near the center of the droplet induces 
two quasiholes of equal charge at the edge.)  

\begin{table}
\begin{tabular}{rccc}\hline\hline
  $N$   &   $E_{\rm CFT}$     &    $E_{\rm CF}$     &  ${\cal O}$\\\hline
  10    &  7.76619(62)    &   7.76600(62)   &  0.9999301(2)  \\
  20    & 24.1403(19)     &  24.1402(19)    &  0.9999274(4)  \\
  30    & 46.3258(18)     &  46.3257(18)    &  0.9999274(3)  \\
  40    & 73.2339(17)     &  73.2339(17)    &  0.9999266(2)  \\
  50    & 102.0588(16)   &  102.0585(16)  &  0.999613(5)   \\
\hline\hline
\end{tabular}
\caption{The Coulomb energy $E_{\rm CFT}$ ($E_{\rm CF}$), quoted in units of 
$e^2/\epsilon\ell$,  for the CFT (composite fermion) two-quasiparticle wave function 
$\Psi_{\rm CFT}$ ($\Psi_{\rm CF}$).  ${\cal O}$ is the properly normalized 
overlap between the two candidate wave functions for a number of system sizes $N$. 
In $\Psi_{\rm CFT}$, we set $L=0$ and $l=1$. The definitions of energy and overlap are given in the text.  The Monte Carlo statistical uncertainty is shown in brackets.}
\label{EnergyAndOverlap}
\end{table}

\subsection{Random Phase Approximation}

The composite fermion quasiparticles have been shown to possess well-defined 
fractional braiding statistics \cite{jain2}. 
The comparisons in the previous subsection imply that this property, in principle,
carries over to the CFT quasiparticles.  However, one of the strengths 
of the CFT description of the FQHE states is that it reveals the braiding 
properties in a transparent manner.  

A calculation of the braiding statistics requires wave functions for 
spatially localized quasiparticles, which are constructed in  the previous section.  A key 
observation is that the braiding statistics can be obtained from the wave function
normalization factor, which depends only on the coordinates of 
localized quasiparticles (c.f. \pref{loc} ). However it is difficult to obtain an explicit analytical expression 
for the normalization factor for localized 
quasiparticle states, because the localized quasiparticle wave functions 
are sums over many correlators; for example, the relevant term in the integral for 
the normalization factor of a single quasiparticle is
\be{singnorm}
\sum_{ij} \langle P_m(z_i)\prod_{k\neq i}V_m(z_k)\rangle 
          \langle P_m(\bar{z}_j)\prod_{l\neq j}V_m(\bar{z}_l)\rangle.\label{NormExample}
\ee
The analytic form of the normalization factor can be obtained by assuming
that only the ``diagonal'' elements in \pref{NormExample} are relevant, which we 
have referred to as the random-phase assumption; the braiding properties of the 
quasiparticles can be derived under this approximation and are in agreement 
with the known results.  In this section we test the random phase assumption  for a single 
localized quasiparticle.

\begin{figure}
\includegraphics[scale=0.63,angle=-90]{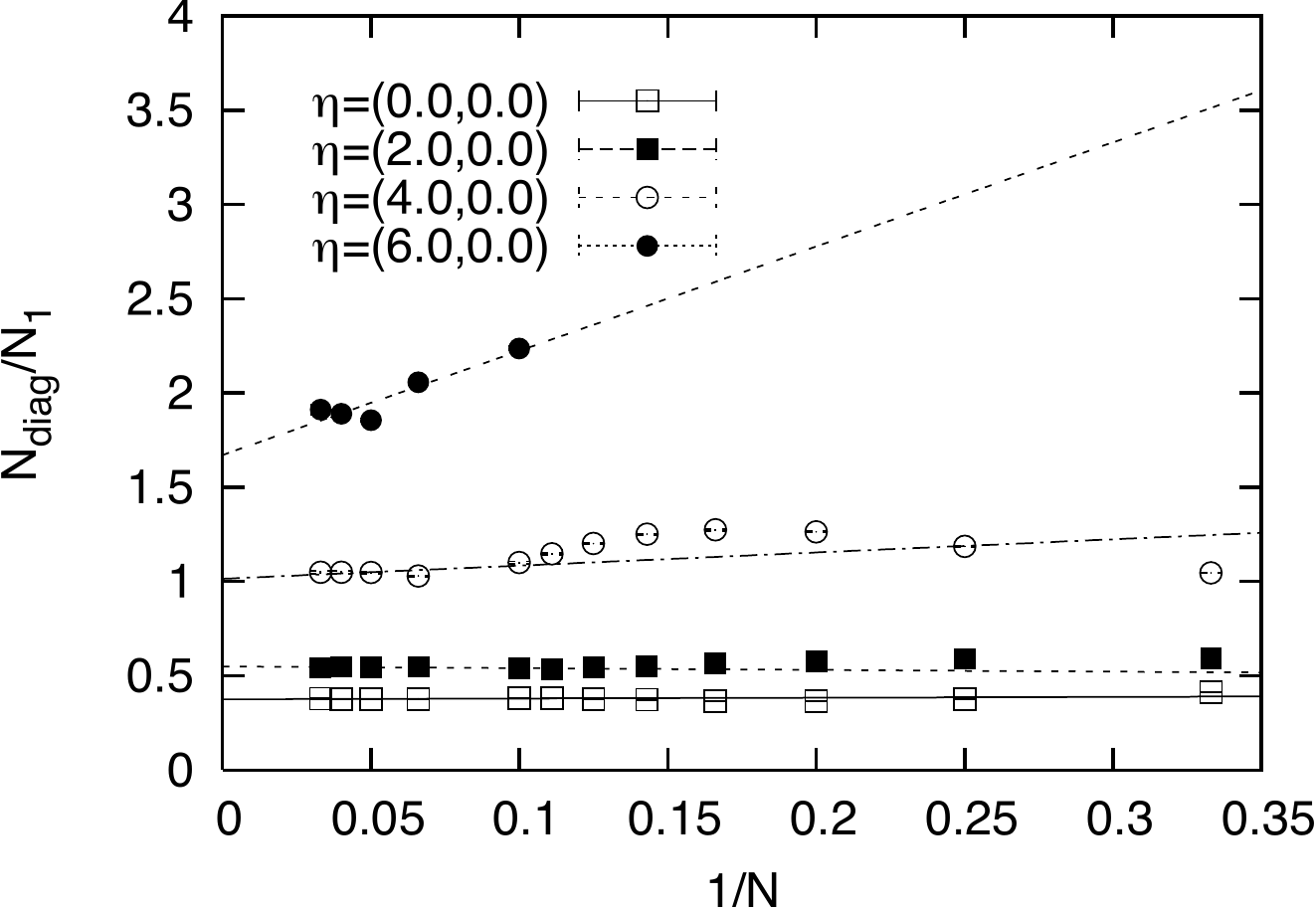}
\caption{The ratio ${\cal M}_{\rm diag}/{\cal N}_1(\eta,\bar{\eta})^2$ for several quasiparticle
location $\eta$. The coordinates are in units of magnetic length.
The results for $\eta=(6.0,0.0)$ are not plotted for $N<10$ because of
significant edge effects.}
\label{RPAplot}
\end{figure}

The wave function for a single quasiparticle at $\nu=1/m$ 
 localized at $\bar{\eta}$ is written as
\be{onewf}
\Psi_{\rm 1qp}(\bar{\eta})=\sum_i(-1)^i\,e^{-\frac{1}{4m}(|\bar{\eta}|^2+2\bar{\eta}z_i)}
\langle P_m(z_i)\prod_{j\neq i}V_m(z_j)\rangle,
\ee
with the correlator $\langle P_m(z_i)\prod_{j\neq i}V_m(z_j)\rangle$ given by 
\be{}
\langle P_m(z_i)\prod_{j\neq i}V_m(z_j)\rangle=
\prod_{j<k}^{(i)}(z_j-z_k)^m\prod_{j\neq i}(z_i-z_j)^{m-1}\sum_{j\neq i}
\frac{m-1}{z_i-z_j}\cdot\exp\left(-\frac{1}{4}\sum_l|z_l|^2\right).
\ee
For simplicity, we define the notation
\be{}
F(z_i)\equiv (-1)^i\,e^{-\frac{1}{2m}(\bar{\eta}z_i)}
\langle P_m(z_i)\prod_{j\neq i}V_m(z_j)\rangle.
\ee
Then the wave function $\Psi_{\rm 1qp}(\bar{\eta})$ can be expressed as
\be{}
\Psi_{\rm 1qp}(\bar{\eta})=e^{-\frac{1}{4m}|\bar{\eta}|^2}
\sum_i\,F(z_i).
\ee
The square of the normalization factor, ${\cal N}_1(\eta,\bar{\eta})$, is given 
by the integral
\ba
{\cal N}_1^2(\eta,\bar{\eta}) 
&=& \int\prod_k dz_k\,\Psi_{\rm 1qp}^*(\bar{\eta})\Psi_{\rm 1qp}(\bar{\eta})\nonumber\\
&=& e^{-\frac{1}{2m}|\bar{\eta}|^2}\int\prod_k dz_k 
\left\{\sum_{i=1}^N |F(z_i)|^2
+\sum_{i=1}^N\sum_{i<j} \left[F^*(z_i)F(z_j)+F(z_i)F^*(z_j)\right]\right\}.\nonumber\\
&\equiv& {\cal M}_{\rm diag}+{\cal M}_{\rm off-diag}.
\ea
${\cal M}_{\rm diag}$ and ${\cal M}_{\rm off-diag}$ are the ``diagonal'' and ``off-diagonal'' contributions
to the full normalization factor, respectively.  We calculate the ratio 
${\cal M}_{\rm diag}/{\cal N}_1(\eta,\bar{\eta})^2$ for several 
quasiparticle locations for $\nu=1/3$. The result is shown in Fig.~\ref{RPAplot}.
The principal conclusion is that the contribution of the single diagonal term is of  
the same order as that of a large number of off-diagonal terms.  Although not conclusive,
this suggests that the diagonal term used to calculate the quasiparticle charge and statistics in section \ref{sec:V} will be dominant, thus providing a partial justification 
for the neglect of the off-diagonal terms.

\section{Summary and Outlook} \label{sec:VII}

\vskip 3mm \noi
In this paper we have extended the class of QH wave functions that can be expressed as correlators in a conformal field theory to include the quasiparticle states at the Laughlin fractions, as well as all the ground states in the positive Jain sequence and their quasiparticle and quasihole excitations. 
The connection between the CFT operators and composite fermions was explicitly demonstrated by constructing $n$ fermionic vertex operators $V_{p,n}$, built from $n$ compactified scalar fields, corresponding to the $n$ filled CF Landau levels at $\nu = n/(2np + 1)$. For these states we also constructed the fractionally charged excitations: 
At $\nu = n/(2np + 1)$, there are $n$ independent hole operators (corresponding to removing a composite fermion from any one of the $n$ filled CF Landau levels) and one unique quasiparticle operator (corresponding to putting a CF in the empty $(n+1)^{st}$ Landau level). The fermionic vertex operator $V_{p,n}$ at level $n$ is closely related to the quasiparticle operator of the Jain state with $n-1$ filled CF LLs; in this sense, the ground state at the fraction $n/(2pn+1)$ of the Jain sequence may be viewed as a condensate of quasiparticles of the state at $(n-1)/[2p(n-1)+1]$.  It should be noted, however, that these quasiparticles obey well defined fractional statistics only in the dilute limit.  We also showed that the conformal field theories used to obtain the CF wave functions give precisely the chiral edge theories that are expected from general considerations within the effective field theory scheme developed by Wen, thus giving microscopic support to that approach. 

An attractive aspect of the methods developed in this paper is that they can be extended and applied to other quantum Hall states. For example, in a recent paper, a straightforward generalization of our vertex operators was employed\cite{BHHK} to describe the states observed recently by Pan {\it et.al.}\cite{pan}, which do not belong to the principal series $\nu=n/(2pn+1)$ but have been modeled as the FQHE of composite fermions \cite{pan,CJ}.  Generalizing our methods to include the negative Jain sequence, $\nu = n/(2np - 1)$, or more generally to states obtained from condensing holes, is a more challenging problem. Another interesting challenge is to find a CFT operator that directly creates a localized quasiparticle, rather than having to construct it as a coherent superposition of angular momentum states as was done in this paper; work on this question is in progress\cite{HHV}.

\vskip 3mm \noi
{\bf Acknowledgements:} We would like to thank J\"urgen Fuchs, Maria Hermanns,  Anders Karlhede and Jon Magne Leinaas for helpful discussions.
We thank Prof. B. Janko for hospitality at the ITS, a joint institute of ANL and the University of Notre Dame, funded through DOE contract W-31-109-ENG-38 and Notre Dame Office of Research. 
This work was supported in part by the Swedish Research Council, by Nordforsk, and by NSF under grant no. DMR-0240458 


\appendix

\section{The background charge} \label{app:A}

\vskip 3mm \noi
Definition of correlators of vertex operators such as \pref{La} requires introduction of a compensating charge to satisfy the neutrality condition implied by the conserved $U(1)$ charge\cite{gula}. This can be done in different ways. The simplest is to put a compensating charge
$
V_{bg}(z_\infty) = e^{-i\sqrt{Nm} \varphi (z_\infty) }
$
at the position $z_\infty$, taken to infinity, and define the correlator
by a limiting procedure:\cite{adon}
\be{altdef}
\Psi_L (z_i)
=\lim_{z_\infty \rightarrow \infty}  z^{mN^2}    \av{ \vm 1 \vm 2 \dots \vm {N-1} \vm N }
=  \prod_{i<j} \jas i j m \, .
\ee
This prescription does not produce the exponential factor
$
 e^{-\sum_i  |z_i|^2/4\ell^2}
$, characteristic of a lowest Landau level wave function.

In this paper we use the prescription given in Ref. \onlinecite{MR}, which corresponds to a smeared background charge given by the operator
\be{bgins}
e^{-i\sqrt m \rho_m \int d^2z\, \varphi(z) } \, ,
\ee
where $\rho_m = \rho_0/m$ with $\rho_0 = eB/2\pi$ the density of filled Landau level.
The difficulty with this prescription is that a direct evaluation of the correlator gives a contribution
\be{diffi}
  e^{- m \rho_m \sum_i \int d^2z\,  \ln (z- z_i)} \, ,
\ee
where the presence of the logarithm makes the imaginary part of the integral undefined.
The aim of this appendix is to give a regularized version of the smeared background charge that: i) is well defined; ii) reproduces the pertinent gaussian factor in \pref{La}; and iii) differs from \pref{La} only through a well defined (although singular) gauge transformation.

The idea behind our regularization  is to replace the continuous background field by a lattice of singular flux tubes of strength
\be{fracflux}
\delta\phi = \frac {\phi_0} n = \frac {2\pi} {ne} ,
\ee
which defines the integer $n$.

First consider a single electron in the presence of a single (fractional) flux tube at the origin. The wave function close to the flux tube behaves as
\be{single}
\psi(z) \sim z^{ -\frac {\delta\phi} {2\pi} } \, ,
 \ee
so for a single electron in a flux tube lattice, it is natural to consider a
wave function of the type
\be{fluxlatt}
\psi(z; \{z_{\vec n}\}) = f(z) \prod_{\vec n} (z - z_{\vec n})^{-\frac {\delta\phi} {2\pi} } ,
\ee
which has the correct singular behaviour at the lattice points $z_{\vec n} $ and satisfies the Laplace equation,
\be{lapl}
\nabla^2 \psi(z; \{z_{\vec n}\})  = 4\partial_z\partial_{\bar z} \psi(z; \{z_{\vec n}\} ) = 0 \ \ \ \ \  ; \ \ \ \ \ z\notin \{z_{\vec n}\}.
\ee
This is not a sufficient condition for an acceptable electron wave function -- we must also require that $\psi(z; \{z_{\vec n}\} )$
is normalizable. As we now show, this will determine the allowed analytic functions $f(z)$.  Without any loss of generality, we specialize to a regular lattice of $K$ points
$z_{\vec n} = (n_x   + i n_y )a $
with spacing $a$ that covers a total area $A$ - this is our regularized version of a uniform flux $\Phi = K\delta\phi$ corresponding to a uniform field of strength $B = \Phi/A$ inside the area $A$. We will ignore edge effects.

Since we have $\lim_{z \rightarrow\infty}  \psi(z; \{z_{\vec n}\} ) \sim f(z) z^{-K\frac {\delta\phi} {2\pi}} $,
and normalizability requires\footnote{
This is the correct condition on a sphere (or a compactified plane). On a real plane the last mode will be marginal in the sense that the normalization integral will have a logarithmic divergence.}
 $\lim_{z \rightarrow\infty}  \psi(z; \{z_{\vec n}\} ) \sim1/z $, we demand $f(z) \sim z^k$ where,
 \be{cond}
 k - K\frac {\delta\phi} {2\pi} = k - \frac \Phi {2\pi} \le -1,
 \ee
 where we choose $A$ so that $\Phi/2\pi$ is an integer. This can be rewritten as
 \be{numblll}
 k + 1\le \frac {\Phi} {2\pi} = A \rho_0 = N_0 ,
 \ee
 where $N_0$ is the number of states in the lowest Landau level.
This result makes it plausible that, in the limit of large $n$, the functions \pref{fluxlatt} with
$$
f(z) = \sum_{k =0}^{N_0 - 1} c_k z^k
$$
 will give a good description of the lowest Landau level at a magnetic field of strength $B$.
 To show this, we rewrite $\psi$ as
 \be{newform}
  \psi(z; \{z_{\vec n}\} ) = f(z)  \prod_{\vec n} \left( \frac { z - z_{\vec n} } { \bar z - \bar z_{\vec n} }     \right)^{-\frac {\delta\phi}{4\pi} }
  \prod_{\vec n}  |z - z_{\vec n} |^{ - \frac {\delta\phi}{2\pi} } \, ,
  \ee
 and approximate
 \be{approx}
   \prod_{\vec n}  |z - z_{\vec n} | ^{ -\frac {\delta\phi}{2\pi} } = \exp\left( -\frac {\delta\phi}{2\pi}   \sum_{\vec n} \ln |z - z_{\vec n} | \right) \approx
   \exp \left( -\frac {\delta\phi}{2\pi}  \frac 1 {a^2}  \int d^2r' \, \ln |\vec r - \vec r' | \right) = e^{- \frac {\delta \phi} {4 a^2} |z|^2 } \, .
   \ee
  The last  integral was calculated as
  \be{intev}
  \int d^2r' \, \ln |\vec r - \vec r' | =\frac 1 4  \int d^2r' \, \ln |\vec r - \vec r' |  \nabla_{r'}^2 r'^2 = \frac {2\pi} 4 \int d^2r' \, \delta^2  (\vec r - \vec r' ) r'^2= \frac {\pi r^2} 2,
  \ee
   where we integrated by parts and neglected boundary terms. (The justification is that there is an understood density function $\rho(\vec r')$ that rapidly falls to zero outside the area $A$ but is essentially constant inside. This still leaves an edge correction due to the derivatives acting on the profile that we ignore.)
Finally we note that $\frac {\delta \phi} {4 a^2} = K\delta\phi \frac 1 {4Ka^2} = \Phi \frac 1 {4A} = \frac {eB} 4 = \frac 1 {4\ell^2}$, where $\ell$ is the magnetic length corresponding to the field strength $B$, so the approximate wave functions are of the form
 \be{appwf}
   \psi(z; \{z_{\vec n}\} ) =  \prod_{\vec n} \left( \frac { z - z_{\vec n} } { \bar z - \bar z_{\vec n} }     \right)^{-\frac {\delta\phi}{4\pi} }
   f(z) e^{-\frac 1 {4\ell^2} |z|^2}
 \ee
 and are expected to be valid in the limit of $a/\ell \rightarrow 0$, and $z$ well inside the lattice.
 We have thus recovered the standard lowest Landau level wave functions, albeit in an unconventional gauge defined by the (arbitrarily chosen) flux lattice.
 This conclusion is also strongly suggested by the numerical calculations of Pryor, who shows that already for $n=8$, at least four flat bands are identifiable in the electron spectrum, corresponding to the four lowest Landau levels\cite{pryor}.
A generalization of the analytic argument presented above to include higher Landau levels would be interesting.

Returning to our original goal of regularizing the operator insertion \pref{bgins}, we see that
\be{regins}
e^{-i\sqrt m \rho_m \int d^2z\, \varphi(z) }  \rightarrow   \prod_{\vec n} V_b( z_{\vec n}  )  = \prod_{\vec n} e^{-i \frac {a^2} {\sqrt m 2\pi \ell^2}  \phi (z_{\vec n} ) }
\ee
will do the job in the limit $ a/\ell^2 \rightarrow 0$, because the total $U(1)$ charge of the $K$ vertex operators $V_b$ equals $-Ka^2 eB/2\pi m = -A\rho_m = -N$, where $N$ is the total number of electrons.
Making the replacement \pref{regins} in \pref{La}
\pref{regins} in \pref{La}
\be{regins-2}
\left\langle V_1(z_1)V_1(z_2)\ldots V_1(z_N)\prod_{\vec n}e^{-i \frac {a^2} {\sqrt m 2\pi \ell^2}  \phi (z_{\vec n} ) }\right\rangle,
\ee
and using the same approximation as in \pref{approx}, we regain the correct exponential factor
(The contraction of $V_1(z_j)$ and $\prod_{\vec n}V_b(z_{\vec n})$
gives the factor $\prod_{\vec n}|z_j-z_{\vec n}|^{-\delta\phi/2\pi}$ which, according to \pref{approx},
gives the Gaussian factor for $z_j$), an unimportant constant from the contractions between the
different $V_b$'s,  and  also
a singular and rapidly changing phase factor just as in \pref{appwf}. This is the regularized version of the statement in reference \onlinecite{MR} that \pref{La} ``is trying to give us the answer in a gauge where the vector potential is zero, which means it differs by an everywhere-singular gauge transformation from the usual symmetric gauge vector potential for the uniform background magnetic field."  Here we should also mention that in correlation functions involving the full scalar field, $\phi(z,\bar z)$, the singular phases will cancel out, so these are well behaved functions even in the limit of vanishing lattice spacing.
Finally, we note that  the regularization procedure outlined above suffers from a formal difficulty - it introduces vertex operators with charges that do not belong to the charge lattice of the CFT under consideration. Putting a compensating charge at infinity would not suffer from this problem, but would also not correspond to a homogeneous system.


\section{Equivalence between CFT and CF wave functions} \label{app:B}
In this appendix we provide derivations for some of the formulae in the main text, and prove that the CFT wave functions for the states in the Jain series indeed reproduce the wave functions from the CF framework using a direct projection on the LLL.  

\subsection{An identity} \label{app:B1}
We begin by deriving the central relation \pref{man}. The basic idea is to express the antisymmetrization as a Slater determinant and then use the Laplace expansion of an $(N+M) \times (N+M)$ determinant,
\be{lap}
\mathrm {det} A = \sum_i  \epsilon_{{\cal P}_i} \mathrm {det} B_i  \mathrm{det} C_i ,
\ee
where the sum is over the $\left(\begin{array}{c} M+N \\ N \end{array}\right)$ ways in which  a $N\times N$ matrix 
$B_i$  can be formed from the first $N$ rows of $A$, $C_i$ is the complementary $M\times M$ matrix and $ \epsilon_{P_i}$ is the sign of the permutation needed to bring the N columns of $B_i$ followed by the $M$ columns of $C_i$ into the original order.\cite{metha}
This formula generalizes the expansion by a row used to derive  \pref{1qp} for the one quasiparticle case. 

We have,
\be{manapp}
& {\cal A} &\{  \prod_{p<q}^M \jas p q {1/m + l_{pq} }    \vtm 1\dots \vtm M \vm {M+1}  \dots  \dots    \vm N  \} \\
 &=& \sum_{{\cal P}_r}   {\epsilon_{{\cal P}_r }}   \prod_{p<q}^M \jas {r_p} {r_q} {1/m + l_{pq} }    
   \vtm {r_1}\dots \vtm {r_M}   \vm { r_{M+1}}  \dots  \dots    \vm {r_N}   \nonumber \\
 &=& \sum_{ \{i_n\} }   {\epsilon_{ \{i_n\}  } }   \sum_{{\cal P}_s} \prod_{p<q}^M {sgn_{{\cal P}_s }}  \jas {s_p} {s_q} {1/m + l_{pq} }    \vtm {s_1}\dots \vtm {s_M} 
 \sum_{{\cal P}_t}  {sgn_{{\cal P}_t }}   \vm {t_{M+1}}\dots \vm {t_N} \nonumber  \\
  &=& \sum_{ \{i_n\} } (-1)^{\sum_{p=1}^M i_p   } {\cal R} \{ \prod_{p<q}^M  \jas {i_p} {i_q} {1/m + l_{pq} }         
   \vtm {i_1}\dots \vtm {i_M}   \vm {\bar i_{M+1}}\dots \vm {\bar i_N}      \} \nonumber \ ,
\ee
where $\{ r_1 \dots r_N\}$ is a permutation, ${\cal P}_r$, of $\{ 1 \dots N \}$;  
$\{ s_1 \dots s_M\}$ is a permutation, ${\cal P}_s$, of a subset $\{i_1 \dots i_M \} $ of the $N$ integers; 
$\{ t_{M+1} \dots t_N\}$ a permutation, ${\cal P}_t$, of the conjugate set $\{\bar i_1 \dots \bar i_M \} $ of $N-M$ integers; and $ {\epsilon_{{\cal P}_r }}  $ \etc the corresponding sign factors.   The symbol $l_{pq}$ denotes the relative angular momentum, and ${\cal R}$ 
is the radial ordering operator.  
The sum $ \sum_{ \{i_n\} }   $  is  over all the 
$N!/M!(N-M)!$ ways of doing this partition,
and $\epsilon_{ \{i_n\} }$ is the sign of the permutation needed to order the $M$ rows of the first partition to the left
in the Slater determinant. 
In the above expression, the second line is the definition, 
the third row follows from the Laplace expansion, and the last by
noting that the prefactor makes the expression explicitly antisymmetric 
under exchange of coordinates in the subset $\{ i_n \}$. Antisymmetry under exchange in the second subset is already
guaranteed by the anti-commutation relations for the $V_1$:s.  As in the single quasiparticle case,  no extra signs are obtained by the final radial ordering because 
$P$ and $V_1$ commute. 
\vskip 3mm
\subsection{Equivalence between the $\nu = 2/5$ CF and CFT wave functions. }
We now prove that the CFT wave function \pref{nqpc2/5} for $\nu = 2/5$ is identical to that of composite fermions. 
Recalling that \pref{nqpc2/5} differs from \pref{nqp} only in that all the derivatives are on the left, we have
the following explicit expression (recall that $N=2M$ so there are $M$ $V_1$:s and $M$ $V_2$:s in
the correlator),  
\be{appnqp}
\Psi_{2/5}^{\rm CFT} (z_i)&=& \sum_{i_1<i_2 \dots i_M} (-1)^{\sum_k^M i_k}  
  \partial_{z_{i_1}} \partial_{z_{i_2}} \dots \partial_{z_{i_M}}   \prod_{k<l}^M  \jas {i_k} {i_l}  { 3}  \\
&& \prs {k_1}  {i_2,i_3\dots i_M}  \jas {k_1}  { i_1}  {2}    
\prs {k_2}  {i_1, i_3\dots i_M}  \jas {k_2}  { i_2}  {2}    \dots
\prs {k_N}  {i_1,i_2\dots i_M}  \jas {k_{N}}  { i_n}  {2}    
 \prjas m n  {i_1, i_2\dots i_M}    3  \, .   \nonumber
\ee
To write this in the CF form, we factor out a full Jastrow factor:
\be{man2}
\Psi_{2/5}^{\rm CFT} (z_i)&=& \sum_{i_1<i_2 \dots i_M} (-1)^{\sum_k^M i_k}  
  \partial_{z_{i_1}} \partial_{z_{i_2}} \dots \partial_{z_{i_M}}   \prod_{k<l}^M  \jas {i_k} {i_l}  {1}
   \prjas m n {i_1, i_2\dots i_M}  { 1}   \prod_{p<q}^{N=2M}  \jas {p} {q}  {2}
\ee
The first two Jastow factors are nothing but the Vandermonde determinants of the
subset $I = \{ z_{i_1} \dots z_{i_M}\}$ and the conjugate subset $J = \{ z_{\bar i_1} \dots z_{\bar i_M}\}$.  Also useful is the operator identity
\be{van}
  \partial_{z_{1}} \partial_{z_{2}} \dots \partial_{z_{M}}  \left|  
\begin{array}{ccccc}
1 & 1& .& .& 1 \\
z_1 & z_2  & .& .& z_M \\
z_1^2 & z_2^2   & .& .& z_M^2  \\
. & .& .& .& . \\
z_1^{M-1} & . & .& .& z_M^{M-1} 
\end{array} 
 \right | =  \left |
 \begin{array}{ccccc}
\partial_{z_1}  & \partial_{z_2} & .& .& \partial_{z_M}  \\
\partial_{z_1}  z_1 & \partial_{z_2}   z_2  & .& .& \partial_{z_M}  z_M \\
\partial_{z_1}  z_1^2 &  \partial_{z_2}   z_2^2   & .& .&\partial_{z_M}  z_M^2  \\
. & .& .& .& . \\
\partial_{z_1}  z_1^{M-1} & . & .& .& \partial_{z_M} z_M^{M-1} 
\end{array} \right |
\ee
which follows because each coordinate, as well as the corresponding derivative, appears once and only once in every term when the determinant is expanded. We can now use the Laplace formula, \pref{lap}, in the opposite direction to write 
\be{man3}
\Psi_{2/5}^{\rm CFT} (z_i) =  \left|  
\begin{array}{ccccc}
\partial_{z_1}  & \partial_{z_2} & .& .& \partial_{z_N}  \\
\partial_{z_1}  z_1 & \partial_{z_2}   z_2  & .& .& \partial_{z_N}  z_N \\
\partial_{z_1}  z_1^2 &  \partial_{z_2}   z_2^2   & .& .&\partial_{z_N}  z_N^2  \\
. & .& .& .& . \\
\partial_{z_1}  z_1^{M-1} & . & .& .& \partial_{z_N} z_N^{M-1} \\
1 & 1& .& .& 1 \\
z_1 & z_2  & .& .& z_N \\
z_1^2 & z_2^2   & .& .& z_N^2  \\
. & .& .& .& . \\
z_1^{M-1} & . & .& .& z_N^{M-1} 
\end{array} \right |  \prod_{p<q}^{N}  \jas {p} {q}  {2} 
= \left |
\begin{array}{ccccc}
\partial_{z_1}  & \partial_{z_2} & .& .& \partial_{z_N}  \\
z_1\partial_{z_1}  &  z_2\partial_{z_2}   & .& .& z_N \partial_{z_N} \\
z_1^2 \partial_{z_1} &  z_2^2  \partial_{z_2}   & .& .&  z_N^2 \partial_{z_N} \\
. & .& .& .& . \\
z_1^{M-1}\partial_{z_1}   & . & .& .& z_N^{M-1} \partial_{z_N}\\
1 & 1& .& .& 1 \\
z_1 & z_2  & .& .& z_N \\
z_1^2 & z_2^2   & .& .& z_N^2  \\
. & .& .& .& . \\
z_1^{M-1} & . & .& .& z_N^{M-1} 
\end{array} \right |  \prod_{p<q}^{N}  \jas {p} {q}  {2},
\ee
where we omitted an unimportant sign. The last expression is, up to an overall normalization factor, the CF wave function as given in reference \cite{Kamilla1997}. 

The last identity in \pref{man3} follows because when the derivatives act on the factors in the determinant they give a row
that is already present in the lower part of the determinant. For this to be true it is necessary that all the angular momentum states are present and that there are at least as many rows without derivatives as those with derivatives. These conditions correspond to having a maximum density droplet of electrons in the second CF Landau level which is no larger then the droplet formed by the electrons in the lowest CF Landau level.  This completes the proof of the statement in the main text.

\subsection{The general CF operators and the Jain series} \label{app:B3}
We now extend the previous analysis to a general state in the Jain series. First we 
give the explicit expressions for the operators $V_{p,n}$ discussed in section \ref{sec:IIID}:
\be{hop}
V_{p,1}(z) &=& e^{i\sqrt{2p   + 1}\varphi_1(z) } \nonumber  \\ 
V_{p,2}(z) &=&  \partial  e^{i\frac {2p} {\sqrt{2p   + 1} } \varphi_1(z) }    e^{i\sqrt{ 1 + \frac {2p} {2p+1}} \varphi_2 (z) }            \nonumber    \\ 
V_{p,3} (z)&=&  \partial^2 e^{i\frac {2p} {\sqrt{2p   + 1} } \varphi_1(z) } 
e^{i\frac {2p} {\sqrt{(2p   + 1)(4p+1)} } \varphi_2(z) }   
 e^{i\sqrt{ 1 + \frac {2p} {4p+1}}   \varphi_3 (z) }                \\ 
 \dots  \nonumber  \\ 
 V_{p,n}(z) &=&   \partial^{n-1} e^{i\frac {2p} {\sqrt{2p   + 1} } \varphi_1(z) } 
 e^{i\frac {2p} {\sqrt{(2p   + 1)(4p+1)} } \varphi_2(z) } 
  \dots 
 e^{i\frac {2p} {\sqrt{[2p(n-2)  + 1][2p(n-1)+1]} } \varphi_{n-1}(z) }   
 e^{i\sqrt{  \frac {2np + 1} {2(n-1)p+1 }}   \varphi_n (z) }    .          \nonumber
\ee
Because all $\varphi_i$'s commute, we can write
\be{simhop}
 V_{p,n} (z)=  \partial^{n-1} e^{i\tilde\varphi_{n}(z) },
 \ee
 where $\tilde\varphi_{1} = \varphi_1$ and 
 \be{effphi}
 \tilde\varphi_{n} (z) =  \sum_{k=1}^{n-1}  \frac {2p}  {\sqrt{[2(k-1)p +1](2kp + 1)} }  \varphi_k  (z) + 
\sqrt{ \frac {2np + 1} {2(n-1)p+1}  }   \varphi_n (z)   \ \ \ \ ; \ \ \ n\ge 0  
 \ee
Using the sum formula:
\be{sum}
\sum_{k=1}^n \frac 1 { [2p(k-1) + 1][2pk +1] } = \frac n {2pn +1} \, 
\ee
and the charge density operator
\be{compcdo}
J(z)   = \frac i {\sqrt {2p+1} }  \partial_z  \varphi_1 (z)  +  \frac i {\sqrt {(2p+1)(4p+1)} }\partial  \varphi_2 (z) \dots
  + \frac i {\sqrt {[2p(n-1)+1][2pn+1]} } \partial \varphi_n (z)
\ee
it can be shown that the operators \pref{hop} satisfy the properties stated in the text vis \'a vis charge and statistics, and also give the filling fraction $\nu = \frac n {2pn +1} $.
We can now construct the wave function for the general ground state in the Jain series by a recursive procedure. For a total of $N=nM$
electrons, it is natural to write
\be{genwf}
\Psi_{p,n}^{CF}  (z_i) &=& {\cal A} \{  \av  { \prod_{i=1}^M V_{p,n} (z_i)   \prod_{j=M+1}^{2M} V_{p,n-1 } (z_j)  \dots
 \prod_{j=(n-1)M+1}^{nM} V_{p,1 } (z_j)  }  \}
\ee
The proof that this indeed reproduces the $\nu = n/(2pn+1)$ CF wave function, is a straightforward generalization
of that given for 2/5 in the previous section. It involves using the Laplace formula \pref{lap} iteratively $n-1$ times, 
breaking the problem down into the $n$ groups (Landau levels) of particles, in analogy with the
procedure in section \ref{app:B1}. The generalization of \pref{appnqp} then contains $n-1$ sign factors, one for
each additional group of particles, and one can follow the logic of \pref{man2} - \pref{man3} (with $n$ $(M\times M)$ subdeterminants
instead of two) to derive the equivalence of the CF and CFT wave functions.


\section{The normalization factors ${\cal N}_1$ and   ${\cal N}_2$    } \label{app:C}

We begin with a single quasiparticle.  Using the explicit form \pref{oneloc}, and keeping only the diagonal terms in the double sum in the normalization integral we get,
\be{onenorm}
|\tilde {\cal N}_1(\eta, \eb)|^{-2} \sim   \sum_{i}  \int d^2 z_i\,    e^{-\frac 1 {2m} |z_i - \eta|^2 }  \int \prod_{j\ne i} d^2z_j \,
\av{ \vtm i   \prod_i \vm i } \, \av{\bvtm i  \prod_i \bvm i }^* \, .
\ee
Here and below we use the sign $\sim$ to indicate that we neglect $\eta$-independent constants.  We write $\vtm i  = \partial_i \hvtm i $ and $\bvtm i = \bar\partial_i \bhvtm i $ and perform the $z_i$ integral after making the approximate substitution $z_i \rightarrow \eta\ $ in the correlators.  This gives
\be{onenormb}
|\tilde {\cal N}_1(\eta, \eb)|^{-2} \sim   \sum_{i}   \int \prod_{j\ne i} d^2 z_j \,    \left[ \partial_{\eta} \av {\hat P_{\frac 1 m} (\eta)    \prod_{j\ne i}  \vm j }\right] \, 
\left[ \partial_{\eb} \av { \hat P_{\frac 1 m}(\eb)   \prod_{j\ne i}   \bvm j }^* \right]
\ee
Note that we first moved the derivatives in the operators $\vtm{}$ outside the expectation values. That this is allowed follows either from a direct calculation, or from noting that $\vtm{}$ is a descendant of the primary field $\hvtm {}$ 
and using standard methods to express the correlator of descendant fields as derivatives of correlators of primary fields\cite{gula}. It is important that all sign factors cancel in the diagonal terms. 

Next we note  that the $\eb$ dependence of each correlator is given by $\av { \hat P_{\frac 1 m} (\eta)    \prod_{j \ne i}  \vm j  } \sim \exp{[-(m - 1)|\eta|^2/(4m)] }$.
This allows us to move the derivatives outside the full two dimensional correlation function.
Reintroducing the magnetic length, $\ell$,  defining  
 $D_\eta = \partial_\eta + c\, \eb $ with $c = (m-1)/(m \ell^2)$, and
 noting $[D_\eta, \bar D_\eta]=0$, we get,
\be{onenormc}
|\tilde {\cal N}_1(\eta, \eb)|^{-2} \sim   \sum_{i}    D_{\eta}  \bar D_\eta   \int \prod_{j\ne i} d^2 z_j \,
| \av {  \hat P_{\frac 1 m} (\eta)    \prod_{j\ne i}  \vm j } |^2  \, 
\ee 
The right hand side of this equation is now in a form where plasma analogy arguments can be applied: the integral is the free energy of an overall neutral  plasma with a charged impurity at the fixed postion $\eta$. 
This free energy is independent of the impurity positions because of screening, so finally, using $ D_{\eta}  \bar D_\eta 1 = c^2\, \eb\eta + c  $, and noting that all terms in the sum give identical contributions, we conclude that to leading order in $\ell^2/|\eta|^2$,  $|\tilde {\cal N}_1(\eta, \eb)|^{-2} \sim \eb\eta $, which gives \pref{berres} in the main text.

The calculation of the two quasiparticle normalization factor, $|\tilde{\cal N}_2(N,\eta; \beb , \eb)|$ follows in an analogous manner, with some extra complication due to the more complicated exponential factors in the expression \pref{loc}.  Again keeping only the diagonal terms and completing squares in the exponents, we get,
\be{denmat2}
| \tilde {\cal N}_2(N,\eta; \beb , \eb)|^{-2} \sim  \frac 1 {\eb\eta} \sum_{i<j} \int d^2\zij d^2 \bzij\, e^{-\frac 1 m|\bzij-N|^2}
 [ e^{-\frac 1 {4m} |\zij - \eta |^2 }
 + e^{- \frac 1 {4m}|\zij + \eta |^2 }  -  2\cos \vartheta \, e^{- \frac 1 {4m} (|\eta |^2 + |\zij |^2    ) } ]   \nonumber  \\ 
( \zij\overline z_{ij})^{1-\frac 1 m}  \int \prod_{k\ne i,j} d^2z_k\, 
[    \partial_i\partial_j  \av {   \hvtm i \hvtm j \prod_{k \ne i,j} \vm k   }  ]     
[   \overline\partial_i\overline\partial_j  \av {   \bhvtm i \bhvtm j \prod_{k\ne i,j} \bvm k   }  ]   ,
\ee
where 
$
e^{i\vartheta} =       \eb\zij - \overline z_{ij} \eta$,  
and overall constants are suppressed and the derivatives are moved outside the expectation values.  
Because of the gaussian factors in $|\bzij-N|$, we can approximate the integral by substituting the maximum value $\bzij= N = \half (\epl + \emi )$.  The third term in the square brackets ($\sim \cos\vartheta $)  is maximum at $\zij=0$; the integral vanishes with this substitution because of the factor $( \zij\overline z_{ij})^{1-\frac 1 m} $.  This term is therefore neglected.  The remaining two terms are equal.  For the first term, 
we have $\zij = \eta = \epl -\emi $.  Proceeding as before,  defining $D_+= \partial_{\epl} + c \,\bar\eta_+ $ {\em etc},
and using  that the $\bar \eta_+ $ dependence of each correlator is $ \sim \exp[{-(m - 1)|\eta|^2/(4m)}] $,  we get, 
 \be{movder}
|  \tilde {\cal N}_2(N,\eta; \beb , \eb)|^{-2} \sim 
  \int \prod_{k\ne i,j} d^2z_k\, [\partial_i\partial_j  \av { \hvtm i \hvtm j \prod_{k \ne i,j} \vm k   }  ]     
[ \bar\partial_i\bar\partial_j  \av { \bhvtm i \bhvtm j \prod_{k\ne i,j} \bvm k   }  ] \\
  =    D_+ D_- \bar D_+ \bar D_-      \int \prod_{k\ne i,j} d^2z_k\, 
| \av { \hvtm i \hvtm j \prod_{k \ne i,j} \vm k   } |^2 .\nonumber
\ee
The integral  is now the partition function for a neutral plasma  with two impurities at positions $\epl$ and $\emi$ and this free energy is again independent of the impurity positions because of screening. We thus have $ D_+D_-\bar D_+ \bar D_- 1  =
c^2[ c^2 |\epl|^2  |\emi|^2 + c|\epl|^2       + c|\emi|^2  + 1]$. 
Finally, taking $N=0$ and substituting   $z_\pm=\pm \eta/2$ in the above expressions, and noting that all terms in the sums give identical contributions, we get, to leading order in $\ell^2/|\eta|^2$,  the formula \pref{berres2} quoted in the text.  Retaining the leading order contribution is valid in the limit when the quasiparticles are far separated; this suggests corrections to statistics for smaller separations, as also found in direct numerical evaluations\cite{kjons,jain2}.

\end{document}